\documentclass[modern, twocolumn]{aastex631}
\usepackage{booktabs}
\usepackage{multirow}
\usepackage{physics}
\usepackage{ulem}
\usepackage{layouts}

\let\tablenum\relax
\usepackage{siunitx}
\DeclareSIUnit\GeV{\giga\electronvolt}
\DeclareSIUnit\TeV{\tera\electronvolt}
\DeclareSIUnit\PeV{\peta\electronvolt}
\DeclareSIUnit\parsec{pc}
\DeclareSIUnit\Gpc{\giga\parsec}
\DeclareSIUnit\Mpc{\mega\parsec}
\DeclareSIUnit\erg{erg}

\NewCommandCopy{\oldeqref}{\eqref}
\renewcommand{\eqref}[1]{Eq.~\oldeqref{#1}}

\newcommand{\E}[0]{\ensuremath{E_\nu}}

\newcommand{\Edet}[0]{\ensuremath{\hat{E}}}

\newcommand{\Epeak}[0]{\ensuremath{E_\mathrm{peak}}}
\newcommand{\Epeakobs}[0]{\ensuremath{\Epeak^\mathrm{obs}}}
\newcommand{\Epeaksource}[0]{\ensuremath{\Epeak^\mathrm{src}}}
\newcommand{\Ebreak}[0]{\ensuremath{E_{0}}}

\newcommand{\Lsrc}[0]{\ensuremath{L^\mathrm{src}}}

\newcommand{\mceq}[0]{\texttt{MCEq}}
\newcommand{\stan}[0]{\texttt{Stan}}

\newcommand{\skyllh}[0]{\texttt{SkyLLH}}

\newcommand{\gammadiff}[0]{\ensuremath{\gamma_\mathrm{d}}}
\newcommand{\gammaps}[0]{\ensuremath{\gamma}}

\newcommand{\Phiatmo}[0]{\ensuremath{\Phi_\mathrm{a}}}

\newcommand{\phidiff}[0]{\ensuremath{\phi_\mathrm{d}}}

\newcommand{\Nex}[0]{\ensuremath{\bar{n}}}

\newcommand{\priorlognorm}[1]{\ensuremath{\mathcal{LN}}(#1)}
\newcommand{\priornorm}[1]{\ensuremath{\mathcal{N}}(#1)}

\newcommand{\txs}[0]{\mbox{TXS 0506+056}}
\newcommand{\fgl}[1]{\mbox{4FGL #1}}
\newcommand{\passoc}[0]{\ensuremath{P_\mathrm{assoc}}}
\newcommand{\hinu}[0]{\texttt{hierarchical\_nu}}
\newcommand{\hnu}[0]{\texttt{hnu}}

\DeclareMathOperator*{\argmax}{arg\,max}

\received{2025 March 6}
\revised{2025 September 15}
\accepted{2025 October 5}
\submitjournal{ApJ}

\shorttitle{Spectral modelling and blazar--$\nu$ associations}
\shortauthors{Kuhlmann \& Capel}

\graphicspath{{./}}

\defcitealias{rodrigues2024spectraicecubeneutrinosin}{R+}

\newcommand{\xaviert}[0]{\citetalias{rodrigues2024spectraicecubeneutrinosin}}

\begin{document}

\title{Impact of multi-messenger spectral modelling on
  blazar--neutrino associations}

\correspondingauthor{Julian Kuhlmann} \email{julian.kuhlmann@mpp.mpg.de}

\author[0009-0004-2166-6909]{Julian Kuhlmann}
\affiliation{Max Planck Institute for Physics \\
  Boltzmannstra\ss e 8, 85748 Garching}
\affiliation{Technical University of Munich \\
  James-Franck-Str. 1, 85748 Garching}

\author[0000-0002-1153-2139]{Francesca Capel}
\affiliation{Max Planck Institute for Physics \\
  Boltzmannstra\ss e 8, 85748 Garching}

\begin{abstract}

  Blazars are interesting source candidates for astrophysical neutrino
  emission. Multi-messenger lepto-hadronic models based on
  proton--photon ($p\gamma$) interactions result in predictions for
  the neutrino spectra (``$p\gamma$ spectra'') which are typically strongly peaked at \si{\PeV}
  energies. In contrast, statistical analyses looking to associate
  blazars and high-energy neutrinos often assume a power-law spectral shape,
  putting the emphasis at lower energies. We aim to examine the impact
  of such spectral modelling assumptions on the associations of
  neutrinos with blazars. We use \hinu{}, a Bayesian framework for
  point source searches, and incorporate the theoretical predictions
  for neutrino spectra through a dedicated spectral model and priors
  on the relevant parameters. Our spectral model is based on recent
  predictions for a selection of intermediate and high synchrotron
  peaked blazars that have been found to be spatially close
  to high-energy events detected by IceCube. We apply our model to the 10 years of
  publicly available muon track IceCube data aimed at point source searches, focusing on the Northern
  hemisphere. Out of 29 source candidates, we find five sources,
  including \txs{}, that have an association probability $\passoc{} > 0.5$
  to at least one event.
  The $p\gamma$ spectra
  typically lead to a lower overall number of associated events
  compared to the power-law case, but retain or even enhance strong associations
  to high-energy events. Our results demonstrate that including more
  information from theoretical predictions can allow for more
  interpretable source--neutrino connections.
 
\end{abstract}


\section{Introduction}
\label{sec:introduction}

Recent years have seen exciting progress in the field of neutrino
astronomy. Since the first observations of astrophysical
neutrinos in the \si{\TeV}--\si{\PeV} range
\citep{Aartsen_2013, Aartsen_2013_science}, the IceCube collaboration,
with increased statistics and improved
event selections, has confirmed the presence of this
diffuse flux of astrophysical neutrinos \citep{2014PhRvL.113j1101A}
(see, e.g.,~\cite{2024PhRvD.110b2001A} and \cite{PhysRevD.107.042005}
for recent measurements of IceCube and Baikal-GVD, respectively).
In our neighbourhood, the Galactic plane has been shown to contribute to the
diffuse neutrino flux \citep{icecube_galactic_plane}. Furthermore, the
data shows evidence for the first sources associated to high-energy neutrinos:
a transient point source, \txs{}
\citep{2018Sci...361.1378I, 2018Sci...361..147I}, and a steady
state point source, NGC~1068 \citep{Aartsen:2020ld, Abbasi:2022sw},
with hints of similar Seyfert galaxies making up the first source
class \citep{abbasi2024icecubesearchneutrinoemission,
  abbasi2024searchneutrinoemissionhard}.

In IceCube searches using predefined source lists, two blazars other
than \txs{} appear to drive incompatibility with the background
hypothesis: \mbox{PKS 1424+240} and \mbox{GB6 J1542+6129}
\citep{icecube_point_source, 2021ApJ...920L..45A,Abbasi:2022sw}%
\footnote{Depending on the study, local significances vary from $2.2\sigma$ to $3.7\sigma$.},
both belonging to the sub-class of ``masquerading BL~Lacs". Despite
exhibiting a nearly featureless spectrum typical of BL~Lacs,
masquerading BL~Lacs have intrinsic line emission that could
contribute to neutrino production by providing an additional
interaction target for energetic protons \citep{2022MNRAS.511.4697P}. In
stacking searches of blazar catalogs only upper limits have been found
by, e.g., ANTARES \citep{2024ApJ...964....3A} and the
IceCube collaboration \citep{2022ApJ...938...38A},
with the flux from blazars corresponding to $\sim1\%$ of the total
diffuse neutrino flux. These results are consistent
with the picture of blazars as transient sources of neutrinos that have a
limited contribution to the observed diffuse astrophysical flux.

Independently of the IceCube collaboration, various associations of
different catalogs containing blazars and IceCube neutrinos have been
found using publicly available information. For example, radio-bright
AGN (\citealt{Plavin:2023gh}, see also \citealt{2023ApJ...954...75A}),
the Roma-BZCAT \citep{Buson:2022ps,
  Buson:2023pf}, and intermediate- and high-synchrotron-peaked BL Lacs
\citep{Giommi_2020,Giommi:2021kd}. The methods used to analyse public
data are typically restricted to spatial correlations between
individual high-energy events or ``hotspots''. In these cases, the
information included in the analysis is limited and therefore further
physical interpretation of the results is challenging. While the
angular distance is a key factor for physical associations, it is also
important to consider the angular resolution of the neutrino telescope,
which quantifies how precisely the detector can reconstruct the
neutrino's direction relative to its true incoming direction (also see
\citealt{2023ApJ...955L..32B}) and event energy information to avoid
spurious results. In particular, for blazars the expected neutrino
emission is typically strongly peaked towards higher energies (e.g.,
\citealt{2014PhRvD..90b3007M,Rodrigues:2024fj}).

In this work, we include these aspects into our analysis and further
examine the physical connection between blazars and neutrinos. We
focus on the impact of using physically motivated neutrino energy
spectra, leveraging the recent results from lepto-hadronic modelling
of BL~Lac blazars \citep[hereafter
\xaviert{}]{rodrigues2024spectraicecubeneutrinosin} that have
previously been proposed as physical counterparts \citep{Giommi_2020}
to IceCube's high-energy alert events\footnote{
The events used in \citet{Giommi_2020} and works building upon this source
and event list include events of IceCat-1, see Sec.~\ref{sec:data_set},
and two further high-energy events included in \citet{2022ApJ...928...50A}.
For brevity we will collectively call these events ``alerts''.} \citep{icecat}. The shape of
these spectra is distinct from the power-law form typically used in
point-source searches, especially at energies
$\lesssim$~\SI{100}{\TeV}. This is due to the modelling of photo-hadronic
interactions presumed to take place in the blazar jets. Energetic protons
may interact with photons from the jet or external radiation fields,
leading to peaked energy fluxes. In contrast, the power laws usually invoked
originate from Fermi acceleration used to explain acceleration of charged particles.

Our analysis is possible thanks to the
\hinu{} \citep{Capel:2024gh}\footnote{\url{https://github.com/cescalara/hierarchical_nu}}
software (hereafter \hnu{}). It is capable
of handling large numbers of free parameters and due to its modular framework
different data sets and detector models. As we have demonstrated in \cite{hnu},
informative prior input is helpful in recovering properties of the sources of neutrinos.
With this approach, we focus on source characterisation rather than background
rejection, quantifying the model and parameters which best represent
the observed data. The resulting posterior distribution also allows us
to evaluate the most likely energies of individual neutrino events and
the corresponding neutrino--source association probability.
With the prospect of more neutrino observatories,
such as KM3NeT \citep{km3net_letter_of_intent} and P-ONE \citep{pone_letter_of_intent},
being online and hence larger data sets, source characterisation will become more important,
and frameworks capable of performing joint analyses will be necessary.
This work is a first step in this direction.

We describe our methods, including the model assumptions, data set and
blazar source selection in Section~\ref{sec:methods}. In
Section~\ref{sec:bayesian_perspective}, we first analyse the
well-studied source \txs{}, demonstrating our ability to reproduce
previous results and quantifying the impact of the different spectral
assumptions. We then extend this analysis to the sample of blazars
from \xaviert{} in Section~\ref{sec:bl_lac_sample}. Finally, we
summarise and conclude in Section~\ref{sec:conclusions}.

\section{Methods}
\label{sec:methods}
In Sec.~\ref{sec:spectral_model}, we introduce the spectral models and prior choices of our analyses.
The statistical formalism is briefly described in Sec.~\ref{sec:statistical_formalism},
and finally we comment on the data selection in Sec.~\ref{sec:data_set} and source selection
in Sec.~\ref{sec:source_selection}.

\subsection{Spectral model}\label{sec:spectral_model}

We use two different spectral models for point sources: a power-law,
for comparisons with results of the IceCube collaboration, and a more
realistic spectrum to test the predicted neutrino emission of
\xaviert{}.

For the power-law spectrum case, we consider
\begin{equation}\label{eq:powerlaw}
  \dv{N}{\E{}} \propto \E^{-\gamma},
\end{equation}
where $\gamma$ is the spectral index and a free parameter and $\E{}$ is
the neutrino energy.

\begin{figure}[!ht]
  \centering
  \includegraphics[width=\columnwidth]{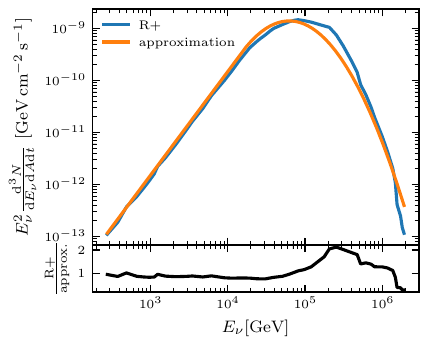}
  \caption{\textbf{Top panel:} Averaged $p\gamma$ spectrum (blue) and
    approximation used in this work (orange). \textbf{Bottom panel:}
    Ratio of $p\gamma$ and approximation.}
  \label{fig:pgamma_spectrum}
\end{figure}

Going beyond this simple description, we model sources based on
results of lepto-hadronic modelling and fits across the
electromagnetic spectrum. For example
we choose the source-averaged neutrino spectrum of Fig.~A.1 in
\xaviert{} (hereafter ``$p\gamma$ spectrum") and approximate it as a
flat spectrum below a break energy, \Ebreak{}, and above as a
logparabola with fixed index of zero and $\beta=0.7$,
\begin{equation}\label{eq:peakyspectrum}
  \dv{N}{\E{}} \propto
  \begin{cases}
    1, & \E{}<E_0\\
    \left (\frac{\E}{E_0}\right )^{-\beta \log{(\E/E_0)}}, & \E\geq E_0.
  \end{cases}
\end{equation}
where \Ebreak{} is left a free parameter. The
parameterisation of \eqref{eq:peakyspectrum} by the peak energy,
\Epeak{}, used in \xaviert{} is obtained by multiplying the
differential flux \eqref{eq:peakyspectrum} with $\E^2$ and finding
the maximum of the resulting expression, leading to
\begin{equation}\label{eq:E0_to_Epeak}
  \Epeak{} = E_0 e^{\frac{1}{0.7}}.
\end{equation}
We display both the numerical prediction and the approximation used,
\eqref{eq:peakyspectrum}, for some arbitrary normalisation and break
energy in Fig.~\ref{fig:pgamma_spectrum}. \xaviert{}
estimated the intrinsic error of
using the source-averaged spectrum rather than the per-source prediction
is on the order of 30\%. This error is further surpassed by the partial insensitivity
of the multi-wavelength fit to varying proton spectra, leading to different
predicted neutrino spectra. In the \hnu{} framework, these uncertainties
are accounted for in a wide prior on the peak energy.

Neutrino emission is assumed to be constant over the time covered
by the analysis for both spectral models considered.

We further model an isotropic diffuse background of astrophysical
neutrinos from unresolved point sources using a power-law,
\eqref{eq:powerlaw}.  The atmospheric neutrino background is modelled
using \mceq{} \citep{fedynitch2015calculation}, employing the hadronic
interaction model \texttt{SIBYLL 2.3c} \citep{Riehn:2017ud} and the
cosmic ray model \texttt{H4a} \citep{Gaisser:2012hf}. The atmospheric
density profile used is \texttt{NRLMSIS-00} \citep{Picone:2002yd}
centered on the IceCube detector and averaged over a year, with the
integrated flux \Phiatmo{} as sole free parameter.  All source spectra
are modelled between \SI{e2}{GeV} and \SI{e9}{GeV} in the detector
frame.  Isotropic source frame luminosities \Lsrc{} are converted to an energy
flux at Earth, assuming a flat $\Lambda$CDM cosmology with
$H_0=\SI{70}{\kilo\meter\per\mega\parsec\per\second},
\Omega_\mathrm{m}=0.7, \Omega_\Lambda=0.3$.
Free high-level model parameters and their prior distributions are
listed in Table~\ref{tab:pl_priors}.  We choose a source-dependent
luminosity prior. For each source we assume a spectral index $\gammaps{}=2.3$ or a source-frame peak energy
$\Epeaksource{}=\SI{e6}{\GeV}$ and compute the according luminosity leading to
0.1 predicted events over the detector lifetime. This luminosity is used as $\mu$ in the lognormal prior,
together with a width of $\sigma=4$.
The value of the spectral index is justified, as $\gammaps{}\sim2$ is expected for blazars.
Given the predicted values of \Epeaksource{} in the \si{\PeV} range,
$\Epeaksource{} = \SI{e6}{\GeV}$ is a reasonable, but not extreme choice.
Due to the luminosity priors being only weakly informative
the exact values do not matter as much.
In Appendix~\ref{app:priors} we have verified that results are robust
against reasonable variations of this choice.
The priors on the astrophysical diffuse spectrum reflect recent
results on the differential flux normalisation \phidiff{} at
\SI{100}{\tera\electronvolt} and spectral index \gammadiff{}
\citep{naab2023measurement}.
We place an uninformative prior on the integrated atmospheric neutrino flux.
The large, $\mathcal{O}(10^3)$, number of atmospheric events included in each fit is constraining
the atmospheric flux normalisation, \Phiatmo{}. Due to a data--Monte Carlo mismatch at low energies,
\Phiatmo{} exhibits an unphysical dependency on parameters of the data selection.
As this issue is restricted to energies below $\sim\si{\TeV}$, it is not affecting the work presented here.
For details we refer to \cite{2025EPJWC.31906004K}.

\begin{table*}
  \centering
  \begin{tabular}[t]{lccc}
    \toprule
    Neutrino Source & Emission model & Param. & Prior \\
    \midrule
    \multirow{5}{*}{Point source} & \multirow{2}{*}{power-law} & $\Lsrc{}$ & $\priorlognorm{\cdot , 4}$ \\
                               & & $\gammaps{}$ & $\priornorm{2.3, 0.5}$ \\
    \cmidrule{2-4}
                               &  \multirow{3}{*}{$p\gamma$} & $\Lsrc{}$ & $\priorlognorm{\cdot, 4}$ \\
                               &  & \multirow{2}{*}{\Epeaksource{}} & $\priorlognorm{\SI{e6}{\GeV}, 3}^{\#1}$ \\
                               &  &                           & $\priorlognorm{\cdot, 0.3}^{\#2}$ \\
    \midrule
    \multirow{2}{*}{Astro.~diff.} & \multirow{2}{*}{power-law} & \phidiff{} & $\priornorm{2.26\times D, 0.20\times D}$ \\
                               &  & \gammadiff{} & $\priornorm{2.52, 0.04}$ \\
    \midrule
    Atmospheric & \texttt{MCEq} & \Phiatmo{} & $\priornorm{0.3\times A, 0.1\times A}$ \\
    \bottomrule
  \end{tabular}
      \caption{Parameters and priors used throughout this work. $\priorlognorm{\mu, \sigma}$ and $\priornorm{\mu, \sigma}$ are lognormal and normal distribution,
      respectively. Units of the diffuse astrophysical flux and atmospheric flux are $D=\SI{e-13}{\per\GeV\per\meter\squared\per\second}$ and $A = \si{\per\meter\squared\per\second}$.
      Prior marked with \#1 is only used in Fit \#1 and defined in the source frame, and replaced by a source-dependent prior in Fit \#2 (marked by \#2), see Sec.~\ref{sec:txs_physical} and \ref{sec:bl_lac_sample} for details.
      The prior of astrophysical diffuse flux is motivated by recent measurements \citep{naab2023measurement}, and of the atmospheric spectrum by the flux necessary to explain the bulk of detected events.
      }
    \label{tab:pl_priors}
  \end{table*}

  \subsection{Statistical formalism}
  \label{sec:statistical_formalism}
  We model emission and detection as a mixture of inhomogeneous
  Poisson point processes, with the mixture components reflecting the
  modelled source components \citep{Streit:2010}. This entails forward
  modelling of all source components to the data space. Further
  difference to the standard method, e.g.,~\citet{Braun:2008bg}
  is that we do not marginalise over the
  neutrino energies but model them as latent parameters. Additionally,
  prior distributions of the high-level parameters are multiplied to
  the likelihood used. The high-dimensional, $\mathcal{O}(10^{3})$,
  parameter space is explored by a Hamiltonian Monte Carlo algorithm
  implemented in the \stan{} programming language \citep{Stan:2024pf},
  returning parameter samples approximating the joint posterior
  density. We summarise marginalised posteriors by ``credible regions'',
  containing $\alpha$ probability. A special case of credible regions are highest posterior
  density regions, which are the narrowest credible regions given a distribution.
  In addition to posterior distributions of model parameters,
  we gain information on an event-by-event basis, finding posterior
  association probabilities of individual events to proposed point sources,
  \passoc{}. For all fits we run 4 chains in parallel with 1000
  warm-up samples and 4000 actual samples per chain to build the target
  distribution. For more details we refer to
  \citet{hnu}.

  As reference results of the frequentist method we find maximum
  likelihood estimates (MLEs) of the expected number of events \Nex{}
  and spectral index \gammaps{} using the publicly available analysis
  tool \skyllh{} \citep{kontrimas2021skyllhframeworkicecubepointsource,Bellenghi:20230u},
  utilising the same public data set, see Sec.~\ref{sec:data_set}.
  It implements the likelihood described in \citet{Braun:2008bg}.
  Despite the different definitions of probability in Bayesian and
  frequentist frameworks, parameter estimates should not be vastly
  different.  Further, evaluating each event's contribution to the
  likelihood at the MLE can be done in a frequentist analysis. The
  most contributing events can thus be found, but, in contrast to the
  Bayesian \passoc{}, the numbers associated lack interpretability.

  \subsection{Data set}
  \label{sec:data_set}
  We apply \hnu{} to 10 years of publicly available IceCube data of
  muon track events aimed at point source searches
  \citep{icecube_data}. Data is selected in a region of interest (ROI)
  of \ang{5} radius around the proposed point source above a
  reconstructed muon energy $\Edet{} = \SI{300}{\GeV}$, yielding up to
  $\sim\num{3700}$ events, depending on the source declination.
  The sources we analyse, see Sec.~\ref{sec:source_selection},
  have been proposed as counterparts to IceCube high-energy alert events.
  IceCube has published a catalog of these events, IceCat-1 \citep{icecat},
  containing a list of all alert events, a most likely neutrino energy assuming
  a power-law spectrum $E_\nu^{-2.19}$, spatial likelihood maps,
  and an effective area as function of declination and neutrino energy.
  For the forward-folding approach used in this work,
  reconstructed muon energies for all events and an energy resolution are needed.
  Hence we have to find the muon track events in \citet{icecube_data}
  corresponding to the alert events.
  Thus we select muon track events within \ang{5} distance to each alert event
  and narrow down possible track counterparts by simultaneous arrival time as the alert
  event, allowing for a difference of \SI{e-4}{\second}. If more than
  one muon track event passes the spatial and temporal selection, we
  choose the highest energy muon track, which typically has an energy
  at least one order of magnitude larger than other muon track events
  with the same arrival time. If no muon track is identified in this manner
  we increase the search radius stepwise up to \ang{10}.
  These steps are needed to quantify the association of the proposed
  source counterparts to the alert events, as the alerts themselves
  are not specified in the public muon track data.
  The angular separation between track and alert best-fitting direction
  are typically  $\sim\ang{0.5}$, but can be as large as $\sim\ang{7}$.
  This is presumably due to the different reconstruction algorithms employed\footnote{
    The 10 year muon track data uses reconstruction algorithms described in \citet{2004NIMPA.524..169A},
    whereas the alert issued in realtime is constructed using methods described in \citet{2017APh....92...30A}.
    The final events published in IceCat-1 are reconstructed scanning over the sky and likelihoods are further
    corrected for systematic uncertainties in, e.g., ice properties and to provide correct coverage \citep{icecat}.
  }.

  \subsection{Source selection}
  \label{sec:source_selection}
  We analyse the sources of
  \xaviert{}, which have been
  found to be spatially close to IceCube alert events.
  The strongest correlation of sources with events was found by
  \citet{Giommi_2020}  when scaling the semi-major and semi-minor axes of the
  90\% uncertainty region of events by a factor of 1.3. Hence
  not all sources will lie within the 90\% uncertainty region
  of the neutrino event which they are suspected to be the
  physical counterpart of. We further
  sub-select only sources in the Northern sky above the celestial
  equator. Below \ang{-5} declination event rates are dominated by
  atmospheric muons. The formulation of the likelihood, in particular
  the forward modelling, necessitates knowledge of the effective area,
  which is not provided for muons, but only neutrinos in the employed data set
  \citep{icecube_data}.

  \section{Application of the method to \txs{}}
  \label{sec:bayesian_perspective}

  First, we present our results for \txs{} using a power-law spectrum in Sec.~\ref{sec:txs_pl}
  to show that we can reproduce the results of previous analyses and
  demonstrate in detail the impact of the spectral shape on our
  results by contrasting the results obtained with the $p\gamma$ spectrum in Sec.~\ref{sec:txs_physical}.

\subsection{Power-law}
\label{sec:txs_pl}
\begin{figure}
  \includegraphics[width=\columnwidth]{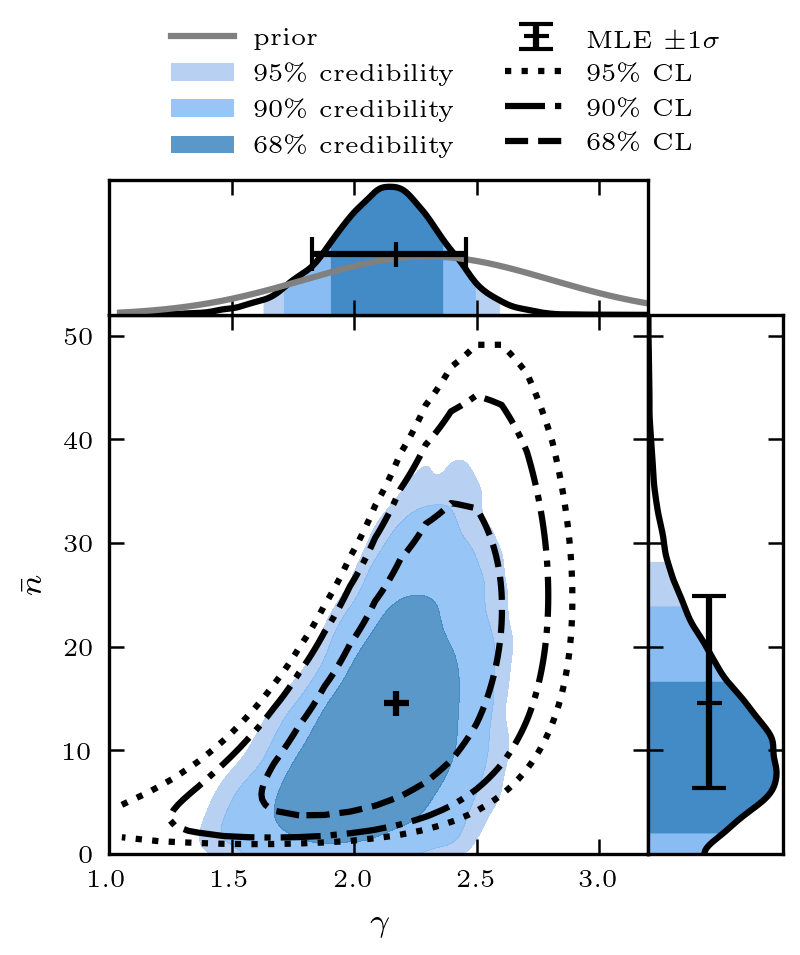}
  \caption{Joint \Nex{} and $\gamma$ posterior. Filled contours show
    68\%, 90\% and 95\% credible regions from darkest to
    lightest. Black contours show 68\%, 90\% and 95\% confidence
    levels found with \skyllh{}. The MLE is marked by a black
    cross. Top and right panel show marginalised posterior densities
    of the same credibilities. Black errorbars show MLE and $1\sigma$
    uncertainties found with \skyllh{}.}
  \label{fig:txs_ps_pdf}
\end{figure}

\begin{figure*}[!ht]
  \includegraphics[width=\textwidth]{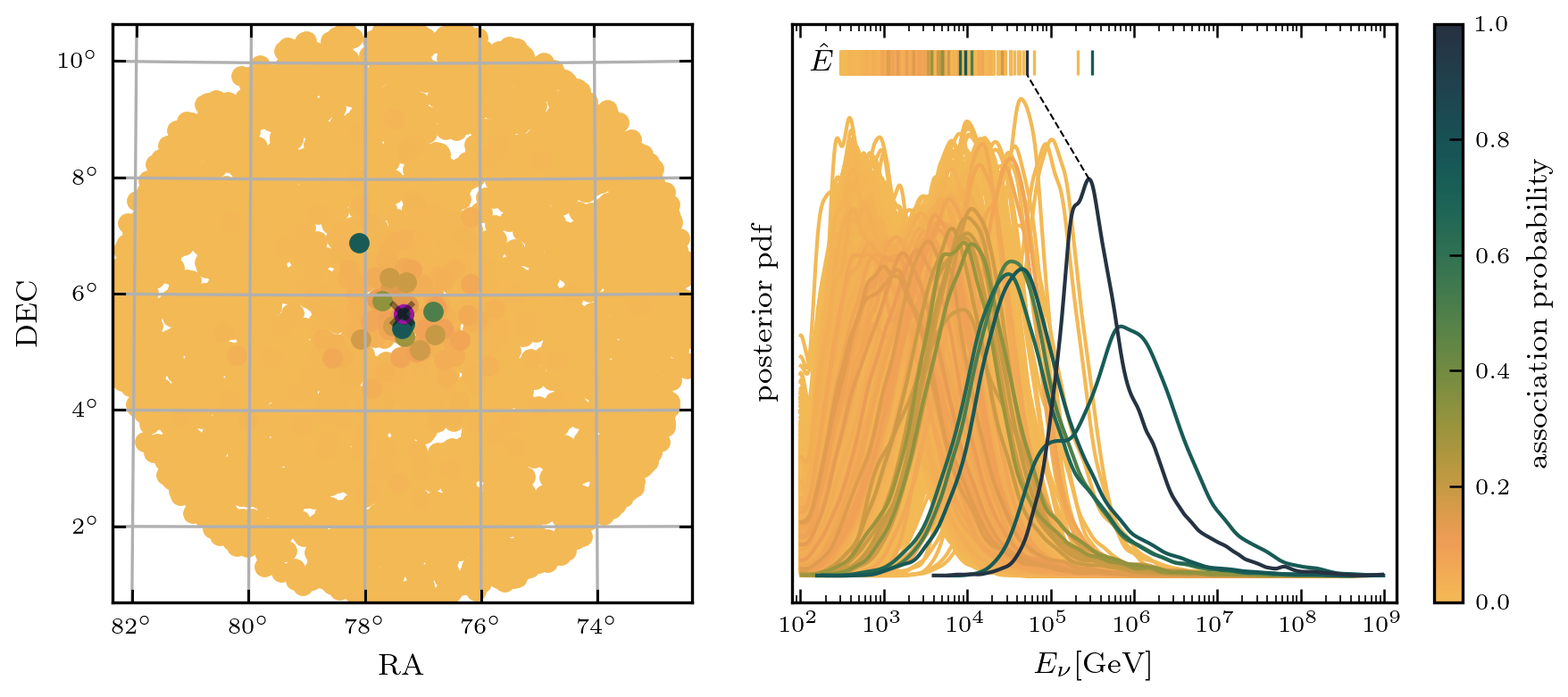}
  \caption{Analysis of \txs{} using a power-law. The colour of dots
    and lines reflects the posterior-averaged event association
    probability to \txs{}.  \textbf{Left panel:} Scatter plots of
    analysed events, projected onto the sky. The dots' size is not
    connected to energy or angular uncertainty. ROI is centered on
    TXS, marked by a cross.  The alert event IC170922A is marked by a
    red circle.  \textbf{Right panel:} Marginalised neutrino energy
    posteriors of all events, transformed to
    $\log_{10}(E_\nu/\si{\GeV})$. The upper axis shows the reconstructed
    muon energy of events, \Edet{}. The reconstructed energy and
    energy posterior of IC170922A are linked by a dashed line.  }
  \label{fig:txs_energy_and_roi}
\end{figure*}

We take \txs{} as test case, using a power-law as the point source neutrino
spectrum.  The point source's joint posterior density of \Nex{} and
\gammaps{} is shown in Fig.~\ref{fig:txs_ps_pdf}, overplotting
confidence levels obtained with \skyllh{}. 
We are able to confirm previous
results of the IceCube collaboration for \txs{}
\citep{2018Sci...361.1378I, 2018Sci...361..147I}.
High-level parameters agree between this work and the
standard frequentist approach. 
We display the individual events' association probabilities
to \txs{} and neutrino energy posteriors in
Fig.~\ref{fig:txs_energy_and_roi}. We see that, in general, high-energy
events nearby the proposed point source have a higher association
probability, consistent with a point source and a hard spectral index,
as indicated by the high-level posteriors.  The alert event IC170922A
has $\passoc{} = 98\%$\footnote{ We reported $\passoc{}=56\%$ for
  IC170922A in a single-event analysis in \citet{Capel:2023Jl}. The
  smaller value can be accounted for by the use of the
  \textit{deposited} energy of \SI{23.7}{\tera\eV} and circularised
  angular uncertainty of the alert, \ang{0.7} \citep{2018Sci...361.1378I}.}.
  We further find an
energy posterior peaked at $\sim\SI{300}{\tera\eV}$, in agreement with
the results of \citet{2018Sci...361.1378I}, and a geometric mean of
$\sim\SI{500}{\tera\eV}$. However, there is a conceptual
difference. The neutrino energy we find has been determined in the
context of fitting a multi-component model to a data selection and
includes all other parameters' uncertainties, rather than determining
properties of an isolated event for a given source spectrum.

\subsection{$p\gamma$ spectrum}
\label{sec:txs_physical}
We now assume the $p\gamma$ neutrino spectrum. In a first fit we use
an uninformative prior (Fit \#1) on the peak energy, \Epeak{}.  In a
second fit, we utilise the information provided by the multi-messenger
fit and place an informative prior on \Epeak{} (Fit \#2). In
particular, we place a source-dependent, narrow log-normal prior on \Epeak{} (see also Table~\ref{tab:pl_priors})
with $\mu$ corresponding to the peak of the energy flux,
i.e.,~$\Epeakobs{} = \argmax{\left(E_\nu^2 \dv{N}{E_\nu}\right)}$ from Table~E.1 in \xaviert{}.

\label{sec:masq_bl_lacs}
\begin{figure*}[!ht]
  \centering
  \includegraphics[width=\textwidth]{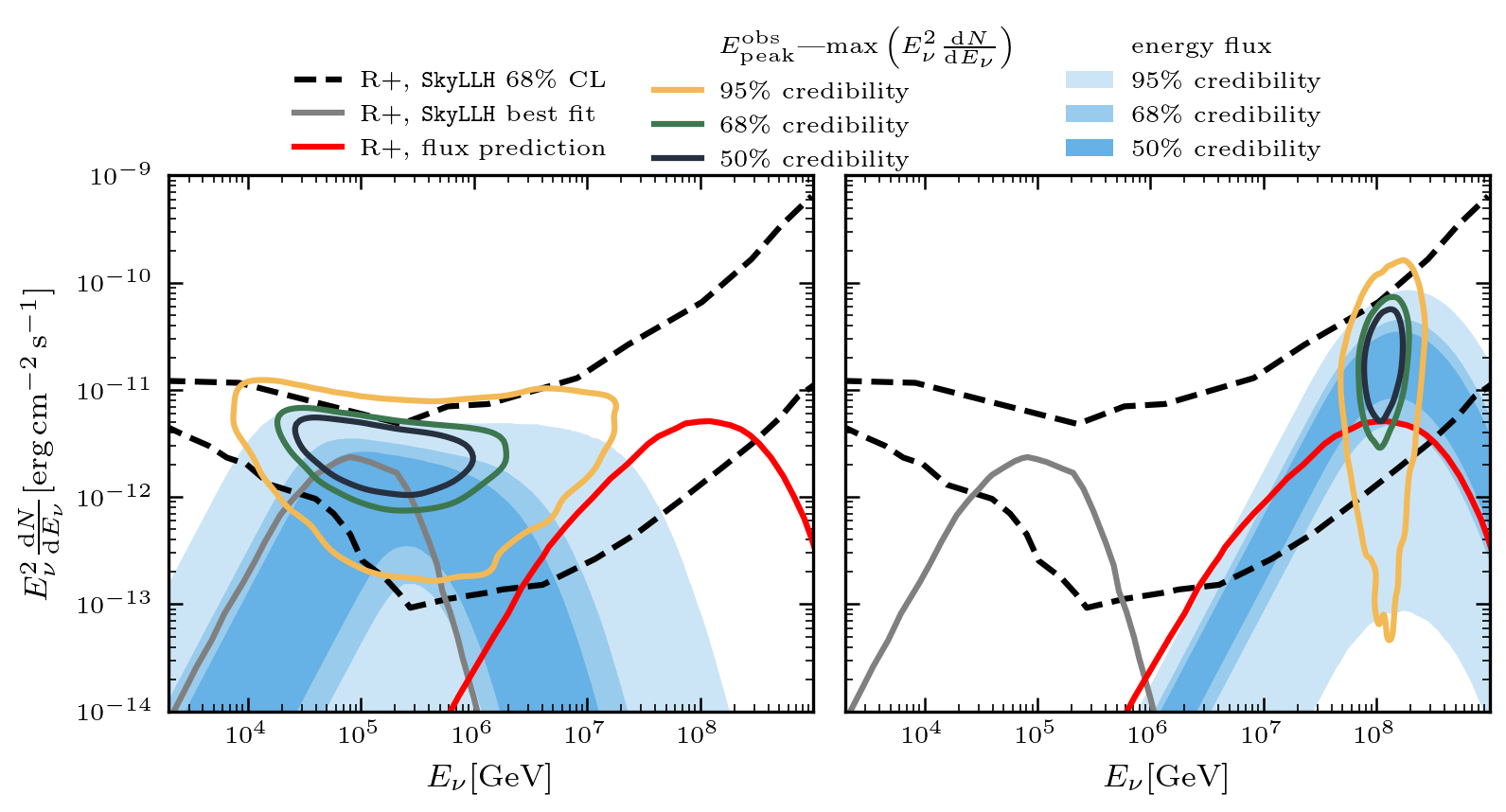}
  \caption{Analysis of \txs{} using the $p\gamma$ spectrum. Coloured
    bands are 50\%, 68\% and 95\% credible regions of fluxes. Dark blue, green
    and yellow closed contours are 50\%, 68\% and 95\% credible regions of
    joint posterior density of \Epeak{} and peak energy flux.  Grey
    line is best fitting neutrino flux found using \skyllh{}, dashed
    contours are 68\% confidence level on \Epeak{} and peak energy flux. Red line is
    neutrino flux prediction.  \textbf{Left~panel:} Uninformative
    priors.  \textbf{Right~panel:} Informative \Epeak{}
    prior. All energies are defined in the detector frame.
    }
  \label{fig:txs_peak_flux_comparison}
\end{figure*}

We compare the resulting energy fluxes against the prediction of
\xaviert{} in Fig.~\ref{fig:txs_peak_flux_comparison}.
The inferred energy flux of Fit \#1, left panel, peaks at energies two to three orders of magnitude lower.
Both predicted \Epeak{} and the predicted flux itself are incompatible at 95\% credibility with our results.
Compared to the frequentist parameter limits at 68\% confidence
(dashed open contours) 
the allowed parameter space of Fit \#1 is much more constrained due to the different definitions
used to set limits on the parameters\footnote{
    The method of \citet{PhysRevD.57.3873} used in \xaviert{} is agnostic about the realised parameters in nature.
    Test statistic distributions for some parameter values of \Epeak{} and \Nex{} are simulated.
    If the simulated distributions are compatible with the test statistic found in the data analysis,
    the corresponding parameters of the simulations are accepted as possibly realised in nature.
}.
We now place a narrow lognormal prior on \Epeak{}, informed by lepto-hadronic modelling of the multi-wavelength observations.
The results of Fit \#2 are shown in right panel of Fig.~\ref{fig:txs_peak_flux_comparison}.
Allowed fluxes are much more constrained compared to Fit \#1, both in the actual flux values and the peak energy.
Compared to the prediction we overestimate the flux by approximately one order of magnitude.
The predicted peak, however, falls into the 68\% credible region.

\begin{figure}[!ht]
    \centering
    \includegraphics[width=\columnwidth]{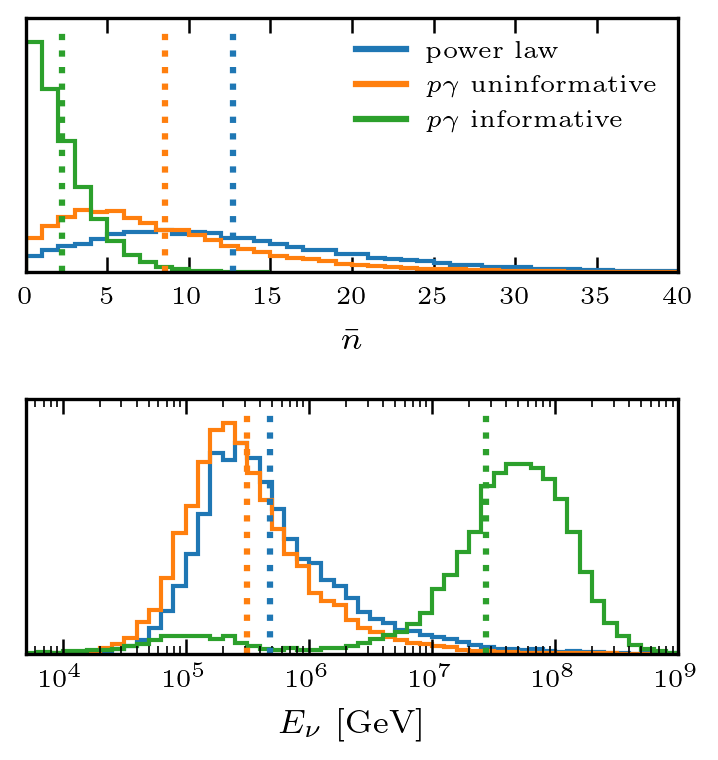}
    \caption{Comparison of fits to \txs{} with different models and priors. \textbf{Top panel:} Posterior of expected number of events. Dashed lines show the posterior means.
    \textbf{Bottom panel:} Energy posterior of IC170922A. Dashed lines show the geometric means of the posteriors.}
    \label{fig:txs_model_comparison}    
\end{figure}

Comparing all three cases in Fig.~\ref{fig:txs_model_comparison}, (power-law, $p\gamma$ with uninformative and informative priors), we find that the number of
point source events decreases in that order. In particular, in the power-law case (Fig.~\ref{fig:txs_energy_and_roi}) a few events at lower energies
around $\Edet{}=\SI{e4}{\GeV}$ have intermediate association probabilities between 30\% and 70\%, which, in part, were detected during the 2014/15 flaring period.
At even lower energies there is a larger number of fainter associations around 10\%.
These events become part of the background due to the informative prior in Fit \#2 (see also second row of Fig.~\ref{fig:app_roi_plots}),
hinting at a strong dependence on the spectral model.
Our prior input is based on models to time-averaged multi-wavelength observations,
while \txs{} exhibited flaring behaviour during the arrival of IC170922A \citep{2018Sci...361.1378I}.
Neutrino emission in 2014/15 also takes place at lower energies \citep{2018Sci...361..147I}.
Reconciling these observations would necessitate time-resolved modelling of
the different flares and quiescent states.

While for both power-law fit and the uninformative $p\gamma$ fit the energy posterior of IC170922A peaks at $\sim \SI{300}{\tera\eV}$,
for the informative $p\gamma$ fit it peaks at $\mathcal{O}(\SI{50}{\peta\eV})$, driven by the informative prior.
For events with intermediate $\passoc$, the employed framework allows for an ``inbetween" case with bimodal energy posteriors.
In the fit to \txs{} using the $p\gamma$ model this is visible in a few events, see second row of Fig.~\ref{fig:app_roi_plots}, right-most panel.
Those events may either be background and have a low parent neutrino energy or
belong to the point source and have an accordingly high neutrino energy.
The density ratio of the posterior modes is consistent with the association probability, for $\passoc{} > 50\%$
the mode at high energies has a higher peak than the low energy mode.

\section{Application to the \mbox{BL Lac sample}}
\label{sec:bl_lac_sample}

We analyse the selected sources, highlighting the impact of different spectral models
in Section~\ref{sec:bl_lac_application}
and determine the associations of high-energy events to their proposed source counterparts in Section~\ref{sec:alert_events}.
\subsection{Fits to public data}
\label{sec:bl_lac_application}
\begin{figure*}[!th]
    \centering
    \includegraphics[width=\textwidth]{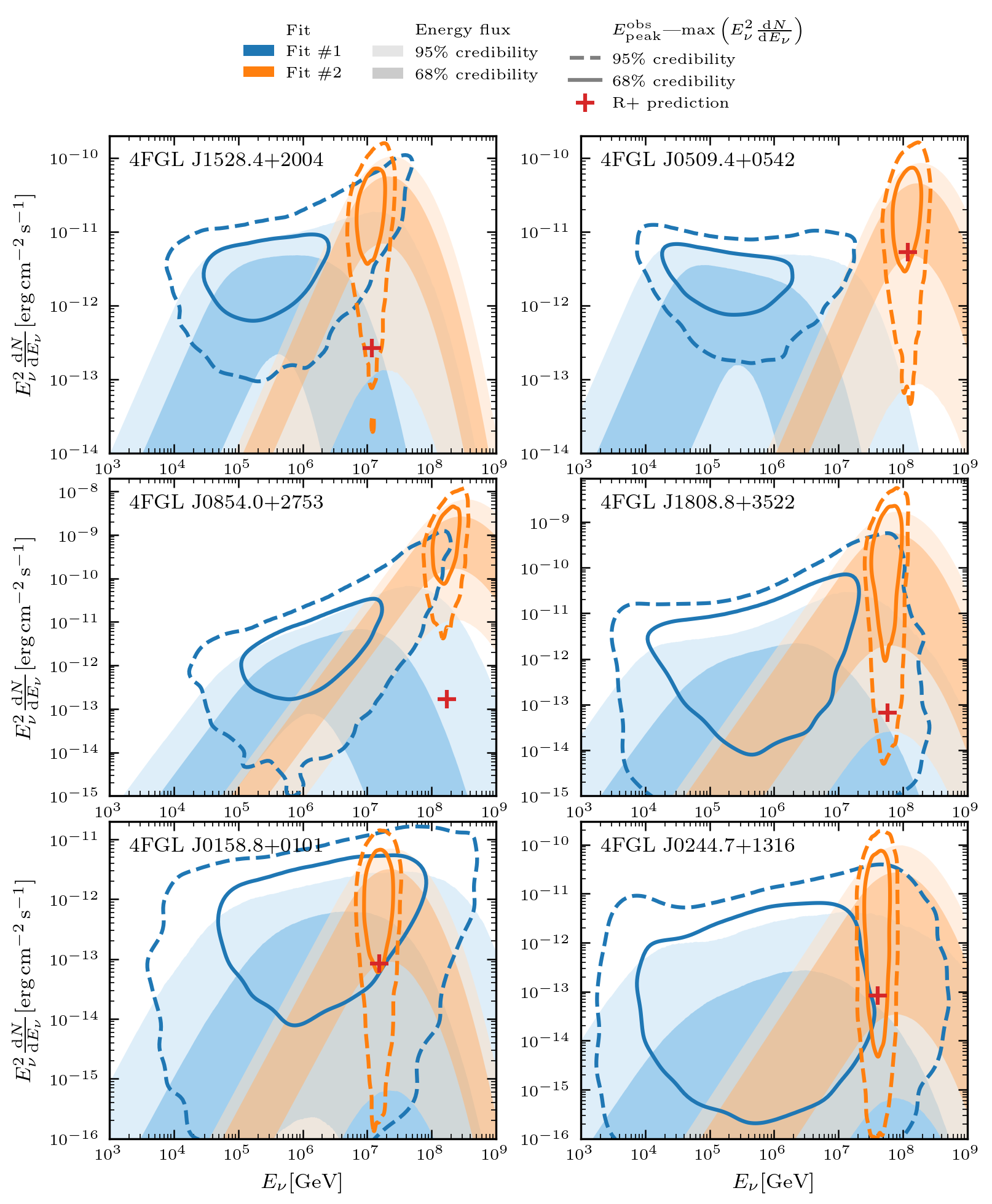}
    \caption{Point source energy fluxes. Filled bands show the 68\% and 95\% credible regions of energy flux
    using an uninformative prior on the peak energy \Epeak{} (blue) and an informative prior (orange). Solid and dashed lines
    show the 68\% and 95\% credible regions of the joint peak energy and peak energy flux posterior.
    }
    \label{fig:flux_compare_prior}
\end{figure*}

\begin{figure*}[!ht]
  \centering
  \includegraphics[width=\textwidth]{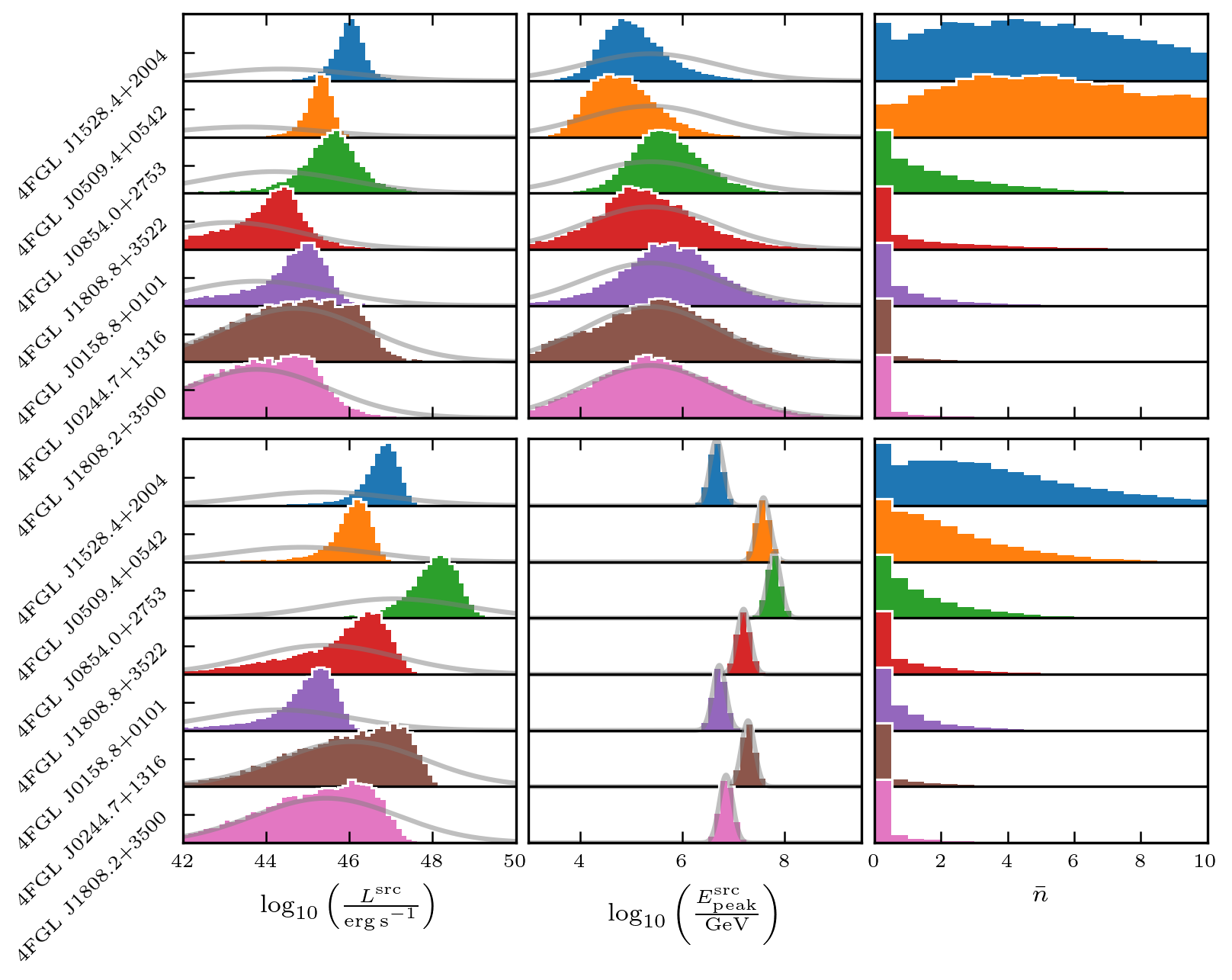}
  \caption{Marginalised posteriors of high-level point source
    parameters. Top (bottom) panels show posteriors obtained with the
    uninformative (informative) \Epeak{} prior. Prior distributions
    are shown in grey, where applicable. \Epeak{} is defined in the source frame.}
  \label{fig:bl_lac_summary}
\end{figure*}

For the entire selection of 29 sources in the Northern Sky we perform three fits:
using a power law, Fit \#1 (uninformative prior) and Fit \#2 (informative prior) using the $p\gamma$ spectrum.
We show results for the top six sources, sorted by \Nex{}, in Fit \#2. Figures including all
sources can be found in a github repository\footnote{\url{https://github.com/specktakel/kuhlmann_capel_pgamma}}.
We show for Fit \#1 and \#2 the inferred energy flux bands in Fig.~\ref{fig:flux_compare_prior},
an overview of the high-level source parameters is presented in Fig.~\ref{fig:bl_lac_summary}.
Plots of the ROIs and energy posteriors are given in Fig.~\ref{fig:app_roi_plots} in the Appendix.

The impact of the prior on the energy fluxes is clearly visible. The additional information constrains the fluxes to smaller enery ranges.
In most cases the flux bands themselves also become narrower.
For the sources for which we can associate neutrino emission, the inferred fluxes of Fit \#2 tend to exceed the predicted peak energy fluxes,
in some cases by multiple orders of magnitude, making a physical association based on the proposed emission mechanism in the source unlikely.
With an informative prior on \Epeak{}, an accordingly high flux is needed to accomodate a high-energy event, if present close to the proposed point source.
For \txs{} the added information leads to a reduction in the number of source events, as already discussed in the previous Section.
In the case of \fgl{J1528.4+2004} the additional prior information on \Epeak{} leads to a \Nex{} posterior less distinct from zero, $P(\Nex{}\geq 1) = 92\%$ decreases to $P(\Nex{}\geq 1) = 83\%$.
We find two events at $\Edet{} = \SI{15}{\tera\eV}, \SI{25}{\tera\eV}$ with high association probabilities of 89\% and 76\% (93\% and 84\% with an uninformative prior).
We can not associate the proposed alert event to \fgl{J1528.4+2004}.
For \fgl{J2227.9+0036} we find, using the uninformative prior, a substantial number of low-energy events with $\passoc{} \geq 5\%$, leading to
$\langle\Nex{}\rangle = 9.0$. All associations vanish and the \Nex{} posterior is consistent with zero when using the informative prior.
We contrast the resulting event associations using a power law and Fit \#2 for this source in Fig.~\ref{fig:app_roi_plots_pl_pgamma}.

We find a further three sources with one event exceeding $\passoc{}=50\%$ in either fit. In all cases this event is the corresponding alert event, see Section~\ref{sec:alert_events}.
For all other sources we report non-detection of emission from the proposed point sources. The \Nex{} posteriors are consistent with zero and the added information on \Epeak{} helps to constrain
source luminosities further, with their posteriors shrinking accordingly.
In general, the posterior-averaged \Nex{} is highest for power laws and further decreases for the informative \Epeak{} prior.
We summarise the resulting \Nex{} for all sources in Tab.~\ref{tab:event_assoc}.

For most sources without evidence for neutrino emissions we find that peak neutrino energy fluxes
are compatible with the predictions of \xaviert{},
in Fit \#1 typically only at 95\% credibility and in Fit \#2 at 68\% credibility.
The \Nex{} posteriors tend to shrink because of the informative priors.
Furthermore, we find stronger constraints on the source luminosities.

The association of many low-energy events at levels of $\sim10\%$ is suppressed by employing the $p\gamma$ model,
and further vanishes in most cases by the use of the informative prior.
Falling power-laws are concentrated at lower energies, whereas the $p\gamma$ spectrum is a flat power-law ($\gamma = 0$) up to the break energy.
Events with high \Edet{} are then driving the position of \Epeak{}. Due to this constraint the
association of lower-energy events to the point source is relatively suppressed. Forcing \Epeak{} to be at high energies via an informative prior
centered on \si{\peta\electronvolt} energies, there is even less chance of observing events of a point source at low (reconstructed)
energies and thus events with a previously low \passoc{} are now firmly part of the background. 
In the case of \txs{}, the number of events with $\passoc{} \geq 5\%$ decreases from 39 to 26 and 5
going from the power-law model to the uninformed $p\gamma$ and finally the informed $p\gamma$ fit.
Accordingly, the energy posteriors of events with considerable association probability are concentrated at very high energies,
$E \gtrsim \SI{10}{\peta\eV}$, e.g.,~Fig.~\ref{fig:txs_model_comparison}.
Compared to the $\sim\SI{300}{\tera\eV}$ found in the power-law analysis of IC170922A, this is a drastic
change in parent neutrino energy.
While this, naively, makes a detection
more difficult, the association of low-energy events appears unphysical in the first place
given the proposed neutrino emission mechanism.

It has been argued that masquerading BL Lacs, due to their line emission acting as a target for proton-gamma collisions, are a promising
source class of high-energy neutrinos, e.g.~\citet{2022MNRAS.511.4697P}. Of the sources analysed in this work showing evidence for neutrino emission or at least
exhibiting a low association to an alert event, we find four masquerading BL Lacs,
\fgl{J0509.4+0542} (\txs{}), \fgl{J1528.4+2004}, \fgl{J0158.8+0101}, and \fgl{J1808.2+3500},
and two true BL Lacs, \fgl{J0854.0+2753} and \fgl{J1808.8+3522}. The latter sources of both lists are linked to the same alert event.

\begin{figure*}
  \includegraphics[width=\textwidth]{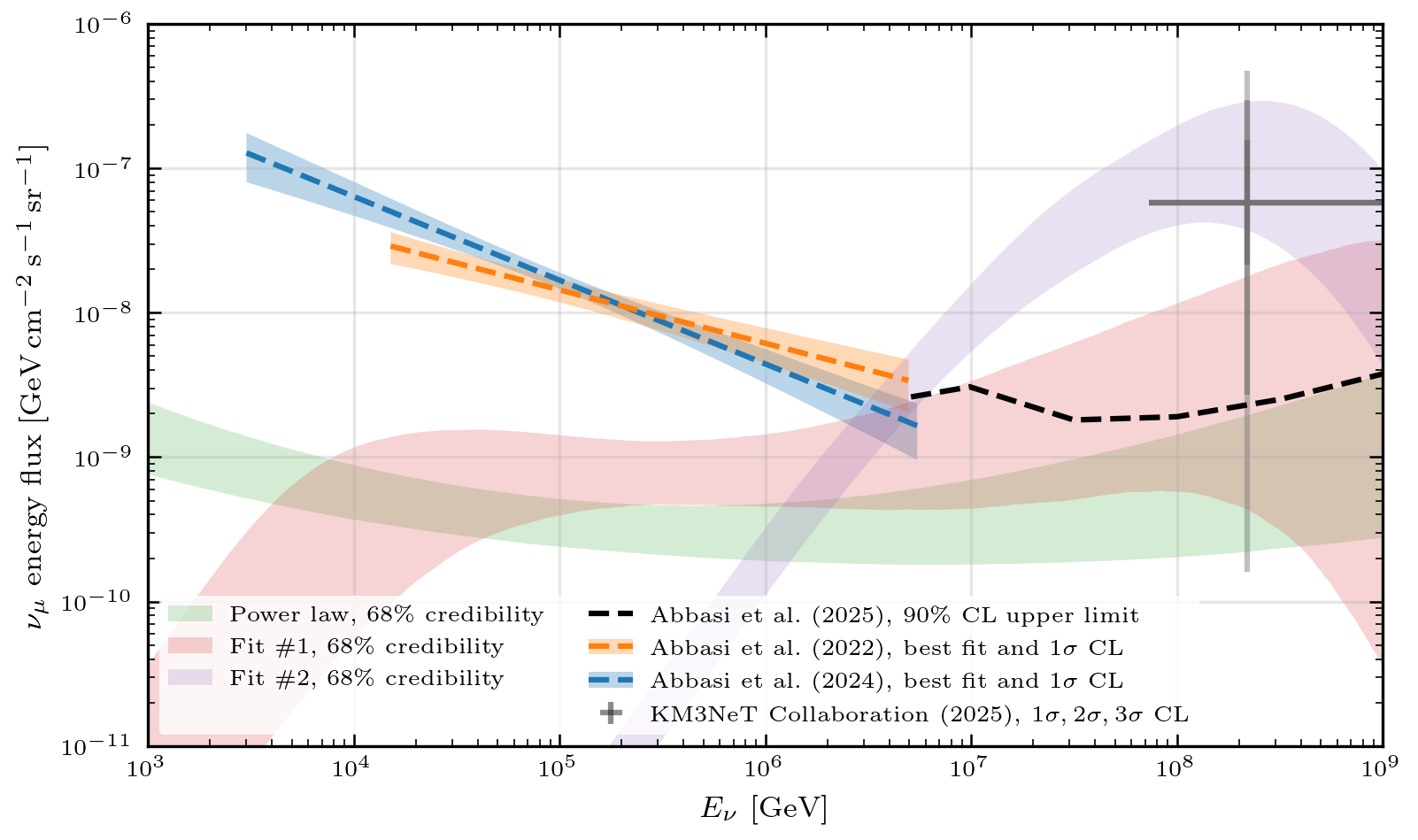}
  \caption{The contribution to the diffuse flux of the 29 blazars
    considered here. Also shown are the diffuse fluxes of \cite{2022ApJ...928...50A}, \citet{2024PhRvD.110b2001A}, \citet[][derived from the KM3--230213A event]{2025Natur.638..376T}, and upper limits of \citet{PhysRevLett.135.031001}.}
  \label{fig:diffuse}
\end{figure*}

We also show our results for the total blazar sample in terms 68\% credibility on their possible contribution to the diffuse flux from the fit to the public muon track data in Fig.~\ref{fig:diffuse}.
Due to the a priori source selection above the celestial equator we normalise the flux by $2\pi$ for the differential \si{\per\steradian}.
For the power law case, our results are well below the diffuse fluxes found by IceCube, e.g., \cite{2024PhRvD.110b2001A}
and agree with the percent-level contribution of blazars to the diffuse flux \citep{2022ApJ...938...38A}.
However, when modelling the contribution of the larger population of blazars to the diffuse astrophysical neutrino flux constraints
can be weaker when employing peaked energy spectra \citep{2016PhRvD..94j3006M}.
Comparisons between different spectral models are non-trivial, as can be seen from our results alone.
For example, the summed flux of Fits \#1 exceeds the flux of the power law fits by a factor of a few at energies above \SI{e5}{\GeV},
despite the lack of informative priors on spectral parameters.
At higher energies, above \SI{100}{\PeV}, both summed fluxes of Fit \#1 (partially) and Fit \#2 exceed the upper limits of \cite{PhysRevLett.135.031001}.
Therein, an $E^{-1}$ power law is used in each decade of neutrino energy as spectral model to derive the differential upper limits.
The summed flux of Fits \#2 is driven by sources at high declination. There, the effective area decreases as the neutrino energy increases,
necessitating an accordingly high flux to account for high-energy events.
In fact, when removing \fgl{J0854.0+2753} and \fgl{J1808.8+3522}, the summed flux
of the remaining sources moves down almost one order of magnitude in both peak energy and peak energy flux.

\subsection{Connection to alert events}
\label{sec:alert_events}
We can link four sources to their respective alert events above $\passoc{} = 50\%$ in either fit.
IC170922A is linked to \txs{} at $92\%$ ($98\%$) probability in Fit \#2 (Fit \#1).
For \fgl{J0854.0+2753}'s alert (IC150904A) we find $\passoc{} = 96\% (93\%)$.
The \Nex{} posterior, while consistent with zero, has a substantial tail to higher values unaffected by the prior choice.
\fgl{J0158.8+0101} is linked to its event (Diffuse (II.1)\footnote{\label{foot:diffuse_ic_names}See \citet[Table 7]{2022ApJ...928...50A}.}) at $\passoc{} = 81\% (69\%)$. 

For the alert event IC110610A two possible counterparts have been proposed, a masquerading BL Lac, \fgl{J1808.2+3500}, 
and a true BL Lac, \fgl{J1808.8+3522}. The latter source displays the higher \passoc{} at 56\% (58\%),
while the former only reaches $22\%$ in Fit \#2.

\fgl{0244.7+1316} can be linked to its alert (IC161103A) at 24\% in Fit \#2.

For all other sources \passoc{} of the linked (alert) event is consistent with zero, notably also for \fgl{1528.4+2004} (Diffuse (II.10)\footnote{See \citet[Table 7]{2022ApJ...928...50A}.}), despite
showing evidence for neutrino emission.
For completeness, we list \passoc{} of the alert events and event with the highest \passoc{} for all sources in Tab.~\ref{tab:event_assoc}.

The best-fitting direction of an alert event and its corresponding event of the 10 year track data typically moves on the order of \ang{0.5}. In case of
\fgl{J1321.9+3219}/IC120515A the separation is approximately \ang{5} and in the most extreme case found is $\sim \ang{7}$.
Additionally, the angular uncertainties are considerably smaller for the corresponding muon tracks in most cases, see Fig.~\ref{fig:app_roi_plots} in the Appendix.
In the case of IC110610A, both proposed counterparts, \fgl{J1808.8+3522} and \fgl{J1808.2+3500}, lie approximately on the 90\% containment radius of the track's angular uncertainty,
and are well within the alert event's 90\% bounding box. Unaccounted for effects have drastic impacts on the results in some cases \citep{kuhlmann_icrc}.
More information is needed in those cases to determine the link between alert event and possible source counterpart.
This can be achieved, for example, by including temporal information in fits, which is planned as a future extension of the \hnu{} framework.

\begin{table*}
    \centering
    \resizebox{\textwidth}{!}{%
      \begin{tabular}{lcccccccccccc}
        \toprule
        \multirow{2}{*}{4FGL} & \multicolumn{3}{c}{Alert event \passoc{}} & \multicolumn{3}{c}{max \passoc{} event} & \multicolumn{3}{c}{$\langle\Nex{}\rangle$} & \multicolumn{3}{c}{$\Pr(\Nex{} \geq 1)$}\\
                              & PL & Fit \#1 & Fit \#2 & PL & Fit \#1 & Fit \#2 & PL &  Fit \#1 & Fit \#2 & PL &  Fit \#1 & Fit \#2 \\
        \midrule
        J1528.4+2004          & 0.00 & 0.00 & 0.00 & 0.87 & 0.93 & 0.89 & 8.6 & 6.6 & 3.8 & 0.88 & 0.92 & 0.83 \\ 
        J0509.4+0542$^{\ast}$ & 0.98 & 0.98 & 0.92 & 0.98 & 0.98 & 0.92 & 12.7 & 8.5 & 2.2 & 0.98 & 0.95 & 0.69 \\ 
        J0854.0+2753$^{\ast}$ & 0.80 & 0.93 & 0.96 & 0.80 & 0.93 & 0.96 & 3.7 & 2.0 & 1.5 & 0.64 & 0.59 & 0.51 \\ 
        J1808.8+3522$^{\ast}$ & 0.43 & 0.58 & 0.56 & 0.43 & 0.58 & 0.56 & 3.9 & 2.1 & 1.1 & 0.47 & 0.44 & 0.35 \\ 
        J0158.8+0101$^{\ast}$ & 0.54 & 0.69 & 0.81 & 0.54 & 0.69 & 0.81 & 1.5 & 1.1 & 0.9 & 0.36 & 0.34 & 0.33 \\ 
        J0244.7+1316$^{\ast}$ & 0.08 & 0.14 & 0.24 & 0.08 & 0.14 & 0.24 & 1.1 & 0.7 & 0.7 & 0.20 & 0.18 & 0.20 \\ 
        J1808.2+3500$^{\ast}$ & 0.11 & 0.19 & 0.22 & 0.11 & 0.19 & 0.22 & 1.4 & 0.8 & 0.5 & 0.22 & 0.19 & 0.16 \\ 
        J2133.1+2529          & 0.00 & 0.00 & 0.00 & 0.17 & 0.24 & 0.04 & 5.0 & 6.1 & 0.5 & 0.40 & 0.41 & 0.11 \\ 
        J1258.4+2123$^{\ddagger}$          & 0.00 & 0.00 & 0.00 & 0.05 & 0.09 & 0.12 & 0.9 & 0.6 & 0.4 & 0.17 & 0.14 & 0.13 \\ 
        J2223.3+0102          & 0.00 & 0.00 & 0.00 & 0.13 & 0.17 & 0.12 & 1.3 & 1.0 & 0.4 & 0.23 & 0.22 & 0.13 \\ 
        J2030.5+2235          & 0.00 & 0.00 & 0.00 & 0.16 & 0.17 & 0.08 & 1.9 & 1.4 & 0.4 & 0.26 & 0.22 & 0.12 \\ 
        J0224.2+1616          & 0.01 & 0.02 & 0.04 & 0.12 & 0.12 & 0.04 & 1.5 & 1.1 & 0.4 & 0.23 & 0.19 & 0.11 \\ 
        J2030.9+1935          & 0.00 & 0.00 & 0.00 & 0.05 & 0.05 & 0.05 & 1.3 & 1.0 & 0.3 & 0.20 & 0.16 & 0.10 \\ 
        J1124.0+2045          & 0.00 & 0.00 & 0.00 & 0.08 & 0.07 & 0.03 & 1.6 & 0.9 & 0.3 & 0.22 & 0.15 & 0.09 \\ 
        J1117.0+2013          & 0.00 & 0.00 & 0.00 & 0.02 & 0.01 & 0.02 & 0.5 & 0.4 & 0.3 & 0.12 & 0.09 & 0.08 \\ 
        J2326.2+0113          & 0.00 & 0.00 & 0.00 & 0.22 & 0.27 & 0.11 & 1.4 & 1.0 & 0.3 & 0.25 & 0.22 & 0.09 \\ 
        J1300.0+1753          & 0.00 & 0.00 & 0.00 & 0.04 & 0.03 & 0.03 & 0.9 & 0.5 & 0.3 & 0.16 & 0.12 & 0.08 \\ 
        J1554.2+2008          & 0.01 & 0.01 & 0.02 & 0.04 & 0.03 & 0.02 & 0.6 & 0.4 & 0.2 & 0.13 & 0.10 & 0.07 \\ 
        J1314.7+2348          & 0.00 & 0.00 & 0.00 & 0.02 & 0.02 & 0.02 & 0.4 & 0.3 & 0.2 & 0.11 & 0.08 & 0.06 \\ 
        J0955.1+3551$^{\dagger}$&  --  &  --  &  --  & 0.03 & 0.02 & 0.02 & 0.7 & 0.4 & 0.2 & 0.15 & 0.08 & 0.06 \\ 
        J1124.9+2143          & 0.00 & 0.00 & 0.00 & 0.01 & 0.02 & 0.02 & 0.4 & 0.3 & 0.2 & 0.10 & 0.08 & 0.06 \\ 
        J1533.2+1855          & 0.00 & 0.00 & 0.00 & 0.04 & 0.06 & 0.01 & 0.7 & 1.2 & 0.2 & 0.14 & 0.30 & 0.06 \\ 
        J2227.9+0036          & 0.00 & 0.00 & 0.00 & 0.34 & 0.47 & 0.03 & 7.4 & 9.0 & 0.2 & 0.54 & 0.64 & 0.06 \\ 
        J0239.5+1326          & 0.00 & 0.00 & 0.00 & 0.01 & 0.02 & 0.03 & 0.3 & 0.3 & 0.2 & 0.08 & 0.07 & 0.06 \\ 
        J1321.9+3219          & 0.00 & 0.00 & 0.00 & 0.03 & 0.02 & 0.01 & 0.5 & 0.7 & 0.2 & 0.12 & 0.17 & 0.05 \\ 
        J0946.2+0104$^{\dagger}$&  --  &  --  &  --  & 0.01 & 0.02 & 0.01 & 0.4 & 0.3 & 0.2 & 0.10 & 0.07 & 0.04 \\ 
        J0232.8+2018          & 0.00 & 0.00 & 0.00 & 0.03 & 0.02 & 0.01 & 0.7 & 0.5 & 0.2 & 0.14 & 0.10 & 0.04 \\ 
        J0344.4+3432          & 0.00 & 0.00 & 0.00 & 0.01 & 0.01 & 0.00 & 0.3 & 0.2 & 0.2 & 0.09 & 0.05 & 0.04 \\ 
        J1003.4+0205$^{\dagger}$&  --  &  --  &  --  & 0.01 & 0.01 & 0.00 & 0.3 & 0.2 & 0.1 & 0.08 & 0.05 & 0.03\\
        \bottomrule
    \end{tabular}%
    }
  \caption{Association probabilities of alert event (columns 2 to 4) and event with the highest source association (columns 5 to 7),
    posterior-averaged number of expected events, $\langle \Nex{} \rangle$ (columns 8 to 10), and probability of $\Nex{} \geq 1$ (columns 11 to 13),
    for all three performed fits (power law, $p\gamma$ with uninformative prior and $p\gamma$ with informative prior).
    For the sources marked with an asterisk the alert event scores highest in all fits.
    For the sources marked with a dagger the associated alert event lies outside the time range covered by the public track data.
    Source marked by a $\ddagger$ appears only in the 3FGL catalog.
}
\label{tab:event_assoc}
\end{table*}

\section{Summary and conclusions}
\label{sec:conclusions}

We have analysed a sample of 29 BL Lacs in the Northern hemisphere taken from \xaviert{},
searching for possible neutrino emission.
In addition to the often-used power-law model for the energy spectrum,
we also implemented a $p\gamma$ spectrum based on the expected
neutrino emission resulting from interactions between matter
and radiation fields present in blazar jets.
Our results show that the choice of energy spectrum has a strong impact
on both the association probability of individual high-energy neutrino alerts
and the overall number of neutrinos that can be connected to the blazars in our sample.

When including an informative prior on the $p\gamma$ spectrum from theoretical predictions,
we see that possible associations with individual energetic neutrinos requires
their true energy to be $\gtrsim$~\SI{10}{\PeV}, roughly two orders of magnitude higher
than that found with a power-law assumption. Furthermore, associations with
lower-energy events that are permitted by power-law fits are no longer viable.
We compare the goodness-of-fit between the power-law and $p\gamma$ cases using
posterior predictive checks and find that both models give a reasonable fit
to the data with no strong preference for one model over the other.

The currently limited public information on IceCube's detector response and
possible systematics could have an impact on the results reported here.
While the energy resolution for through-going track events used here is
naturally limited due to the unknown location of the interaction point of the neutrino,
the coarse binning of the public energy resolution (in both neutrino energies and declination)
exaggerates this effect, making it harder to distinguish possible parent neutrino energies.
For studies on the different results of IceCube-proprietary and public data we refer
to \citet{Bellenghi:20230u,chiara_bellenghi_dissertation}.
Furthermore, the event--source association strongly depends on the angular separation
of event and source and angular resolution used. We note that there can be significant differences
in the best-fit direction reported for the alert events and the corresponding
cross-matched event in the public track dataset.
In extreme cases, this systematic shift can be larger than the reported
reconstruction uncertainty. Additionally, the 2-dimensional Gaussian distribution used
to describe the angular resolution may not be the best description of
the reconstructed event direction and has been updated in the latest IceCube analyses
(e.g.~\citet{Abbasi:2022sw} and the supplementary material).
However, we do not expect these considerations
to impact the overall conclusions of our work regarding the importance of
spectral modelling in determining source--neutrino associations.

Considering the physically-motivated and informative $p\gamma$ model,
we find four promising blazar source candidates (in addition to \txs{})
that still show a strong connection to neutrino events in this more constraining framework:
4FGL J1528.4+2004, 4FGL J0854.0+2753, 4FGL J1808.8+3522 and 4FGL J0158.8+0101.
Two of these sources are masquerading Bl~Lacs and two are true BL Lacs.
These sources have $\passoc{} > 0.5$ to individual energetic events
and an expected number of neutrino events of $\bar{n} \gtrsim 1$.
Given that blazars are highly variable sources, further analysis making use of
time-dependent spectral modelling will be necessary to determine if these
associations make physical sense. While this is challenging due to
the availability of simultaneous observations across multiple wavelengths,
we plan to investigate this direction in future work.

We further see the impact of the power-law and $p\gamma$ spectral assumptions
in our results for the constraints on the diffuse flux from all the blazars
in our sample. As expected, the $p\gamma$ spectrum leads to a much larger
contribution in the 10--\SI{100}{\PeV} range while still being consistent
with the public IceCube tracks data set considered here.
These results are particularly interesting in light of the recent detection
of a $\sim$~\SI{100}{\PeV} neutrino event by KM3Net \citep{2025Natur.638..376T}.
While there are several blazars consistent with the uncertainty region,
thus far there is no conclusive evidence for an association \citep{KM3Net:2025ps},
and it remains unclear whether this event is of cosmogenic origin.

IceCube has announced an updated catalog of alert events, IceCat-2, with improved
reconstruction uncertainties \citep{2025arXiv250706176Z}. As we only use events of
the 10 year public track data in our analyses, the improved reconstruction will only help
indirectly by providing a better handle on angular systematic uncertainties
through crossmatching events,
which will help to infer the true relation between source candidates
and high-energy neutrinos \citep{kuhlmann_icrc}.

Our results demonstrate the importance of physical spectral modelling
for source--neutrino associations. In future, we plan to include
complementary data sets with improved energy resolution (e.g.~cascades, starting events)
and time-dependent information into our analyses to improve the constraining power
and further enable the interpretable connection of neutrinos with candidate sources.

\begin{acknowledgements}
  The authors express their thanks to Xavier Rodrigues and Martina
  Karl for discussions about the multi-wavelength modelling,
  and to Cristina Lagunas Gualda for discussions about
  IceCube's alert events and feedback on the draft. The authors further acknowledge 
  Thara Caba for crossmatching alert events with the public track data, and
  Jarred Green for providing the colour map used in this work.
  J.K.~acknowledges support from the DFG through the
  Sonderforschungsbereich SFB 1258 ``Neutrinos and Dark Matter in
  Astro- and Particle Physics" (NDM).
\end{acknowledgements}

\vspace{5mm}
\facilities{
    Computations were performed on the HPC systems Raven and Viper at the Max Planck Computing and Data Facility.
    This research has made use of the SIMBAD database, operated at CDS, Strasbourg, France \cite{simbad}.
    }

\software{
    \texttt{arviz} \citep{Kumar_ArviZ_a_unified},
    \texttt{astropy} \citep{2013A&A...558A..33A,2018AJ....156..123A},
    \texttt{h5py} \citep{collette_python_hdf5_2014,h5py_7560547},
    \texttt{matplotlib} \citep{Hunter:2007},
    \texttt{numpy} \citep{numpy},
    \texttt{scipy} \citep{2020SciPy-NMeth},
    \texttt{seaborn} \citep{Waskom2021},
    \texttt{stan} and \texttt{cmdstanpy} \citep{Stan:2024pf}.
}

\appendix
\restartappendixnumbering

\section{Result plots}
\begin{figure}[!ht]
    \centering
    \includegraphics[width=\textwidth]{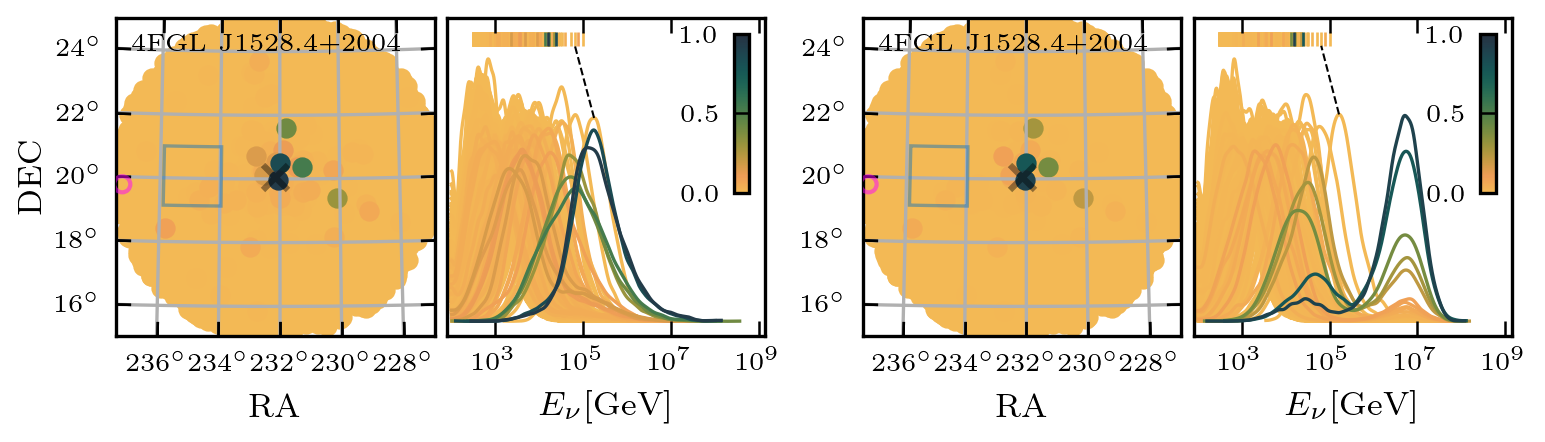}
    \vspace{-1cm}
    \includegraphics[width=\textwidth]{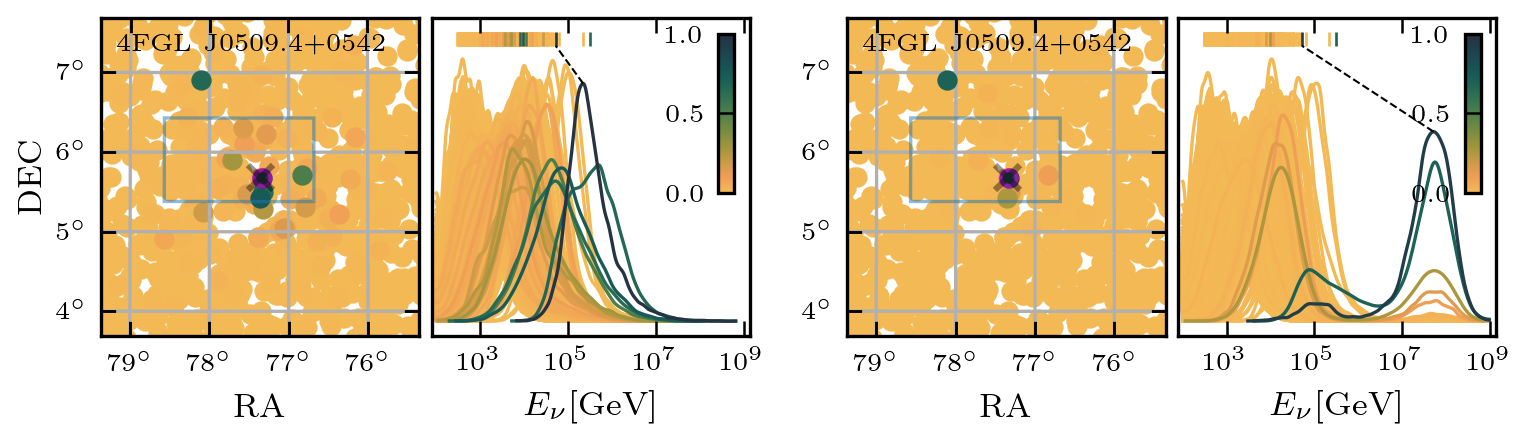}
    \includegraphics[width=\textwidth]{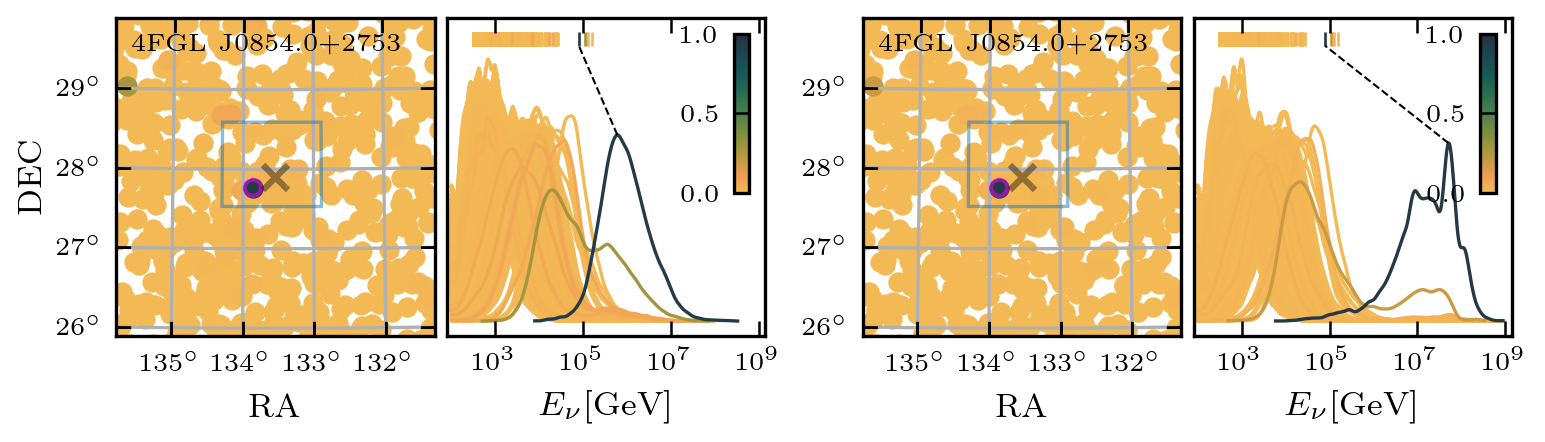}
    \includegraphics[width=\textwidth]{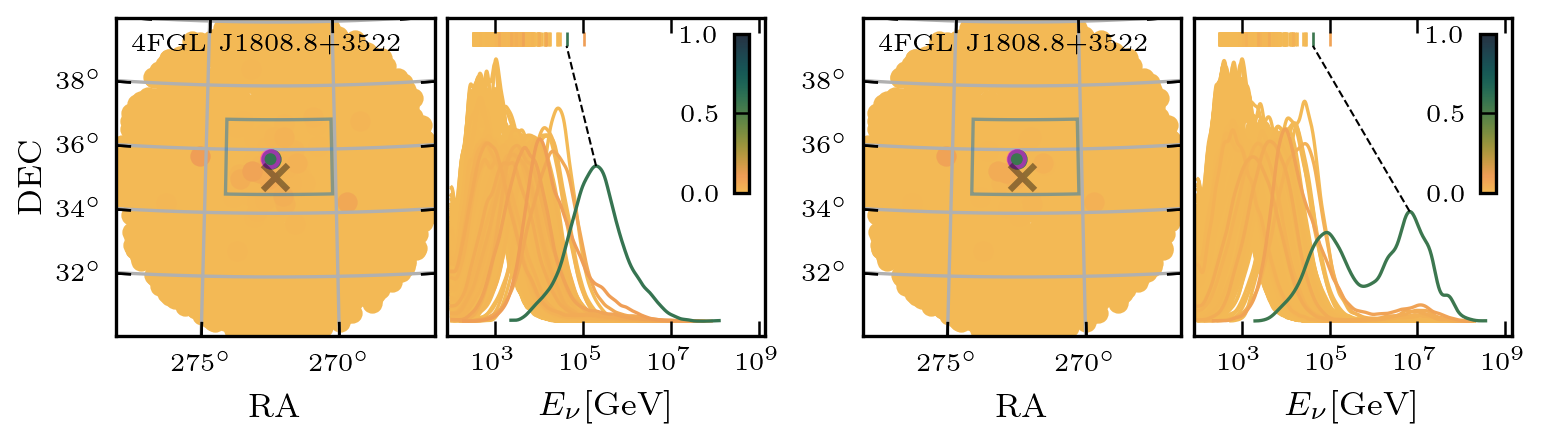}
    \caption{Event-wise analyses of fits. Left (right) column shows results using an uninformative (informative) \Ebreak{} prior.
    Position of alert events included in \citet{icecube_data} are marked by a magenta circle, their 90\% angular uncertainty is shown as a grey circle.
    A box enclosing the 90\% likelihood contour of associated IceCat-1 events is plotted in light blue.
    The alert events' reconstructed energy is linked to their respective energy posterior by a dashed line.}
    \label{fig:app_roi_plots}
  \end{figure}

We show event-by-event analyses for the top six sources, sorted by \Nex{},
in Fig.~\ref{fig:app_roi_plots}, contrasting the results of Fit \#1 and \#2.
In Fig.~\ref{fig:app_roi_plots_pl_pgamma}, we highlight the reduction of low
to intermediate event associations by comparing fits using power laws and the $p\gamma$
model with an informative prior.
  
  \renewcommand{\thefigure}{A\arabic{figure} (Cont.)}
  \addtocounter{figure}{-1}
  
  \begin{figure}[!ht]
    \centering
    \includegraphics[width=\textwidth]{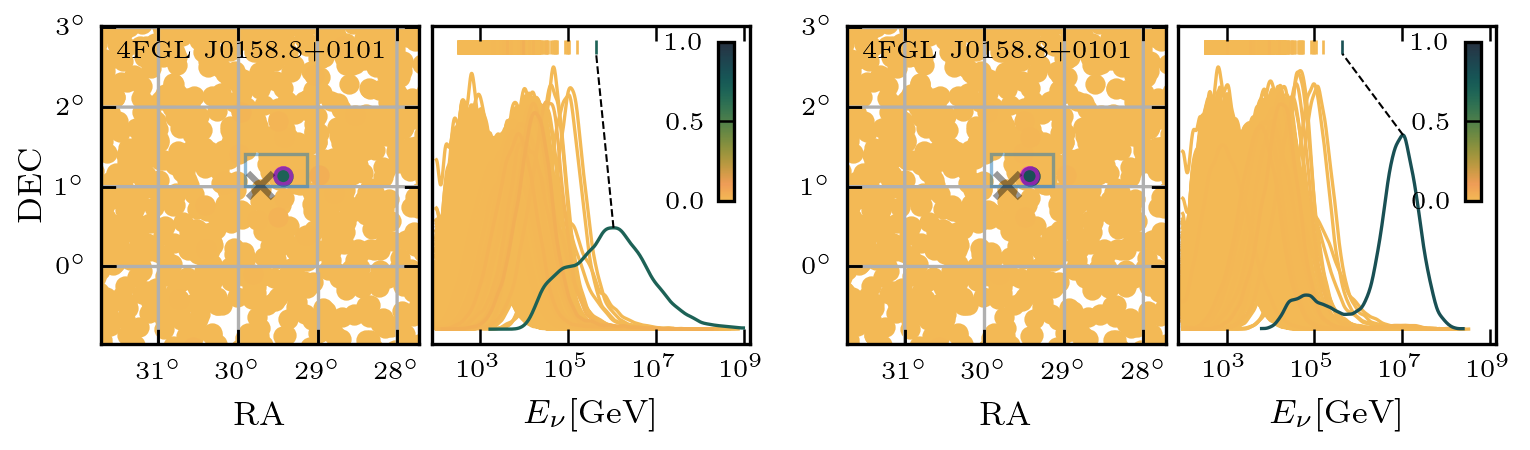}
    \includegraphics[width=\textwidth]{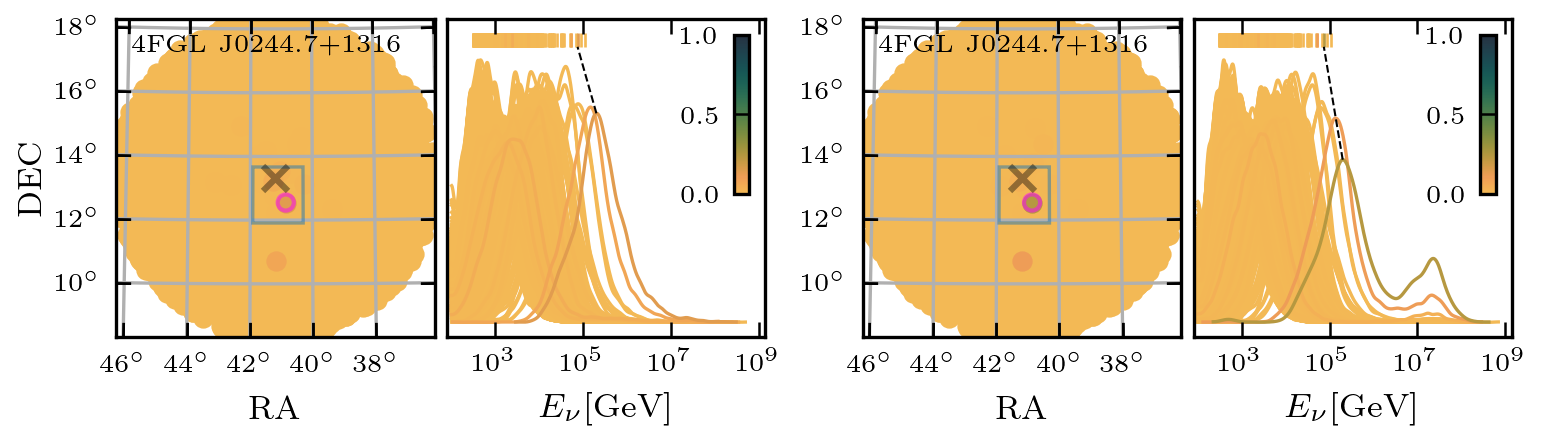}
    \includegraphics[width=\textwidth]{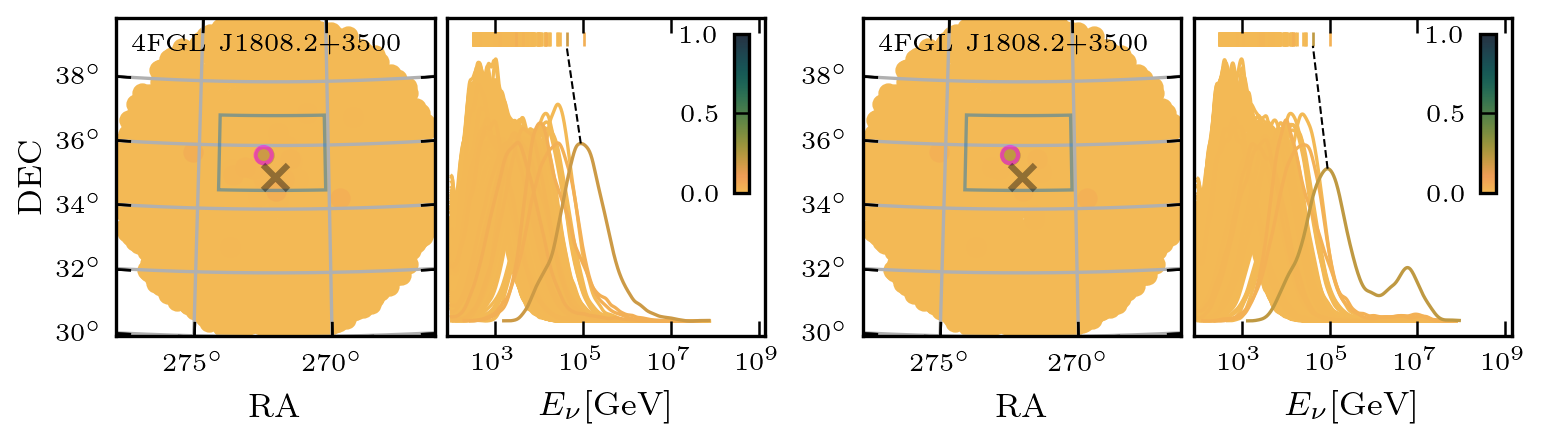}
    \caption{}
    \label{fig:app_roi_plots_cont}
\end{figure}
\renewcommand{\thefigure}{A\arabic{figure}}

\begin{figure}[!ht]
  \centering
  \includegraphics[width=\textwidth]{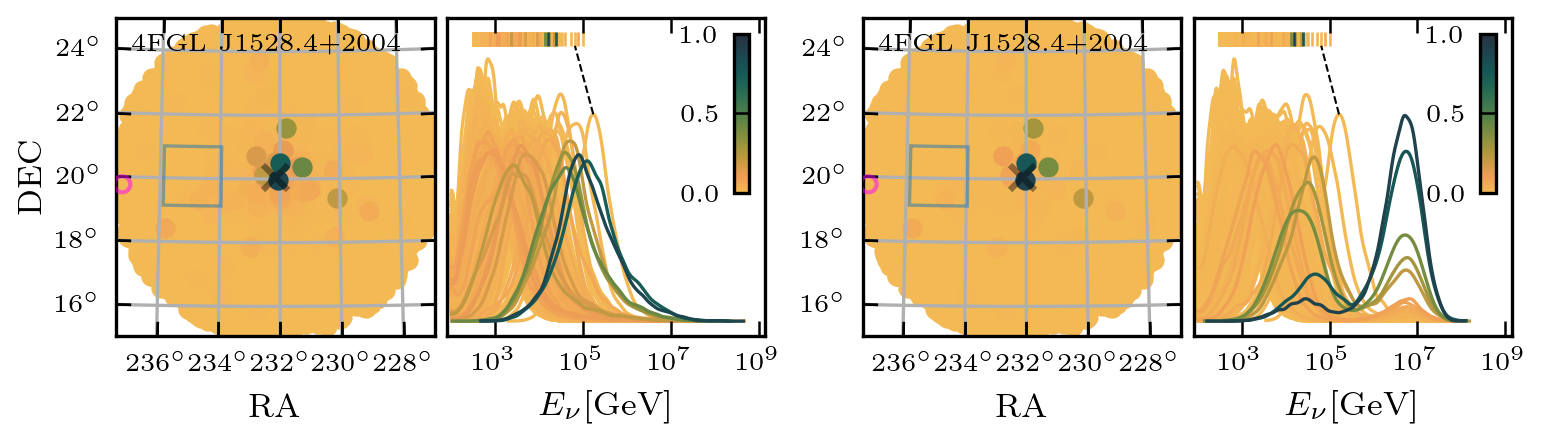}
  \includegraphics[width=\textwidth]{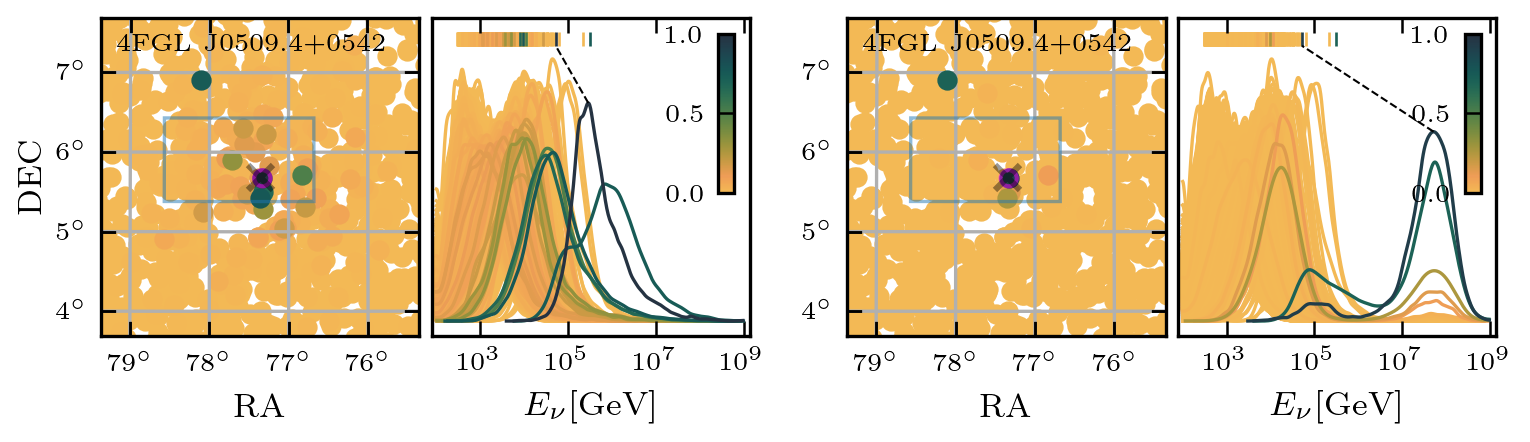}
  \includegraphics[width=\textwidth]{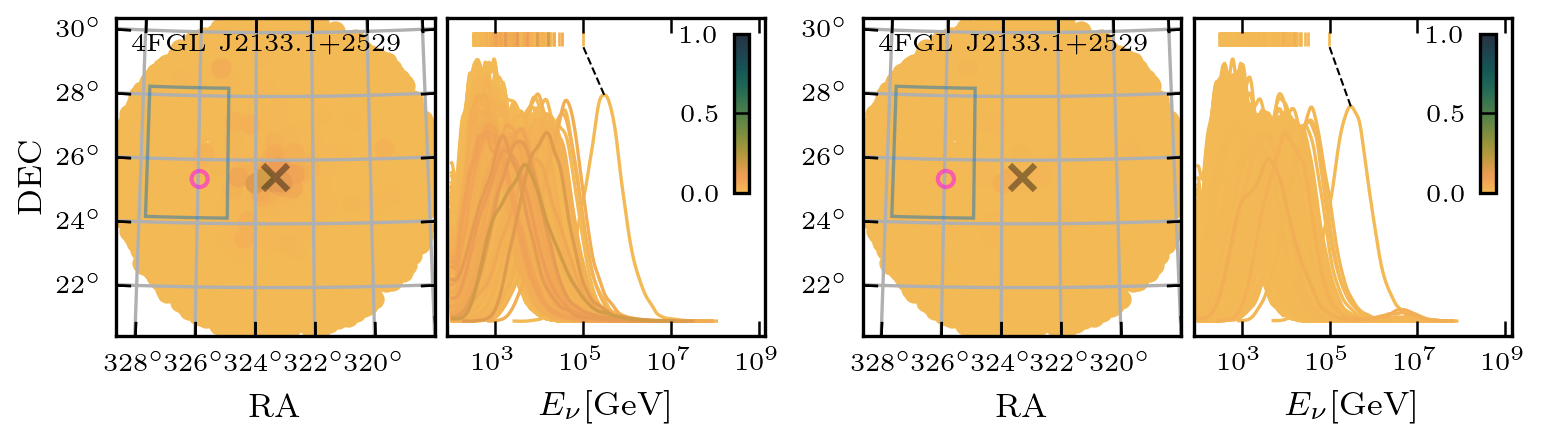}
  \includegraphics[width=\textwidth]{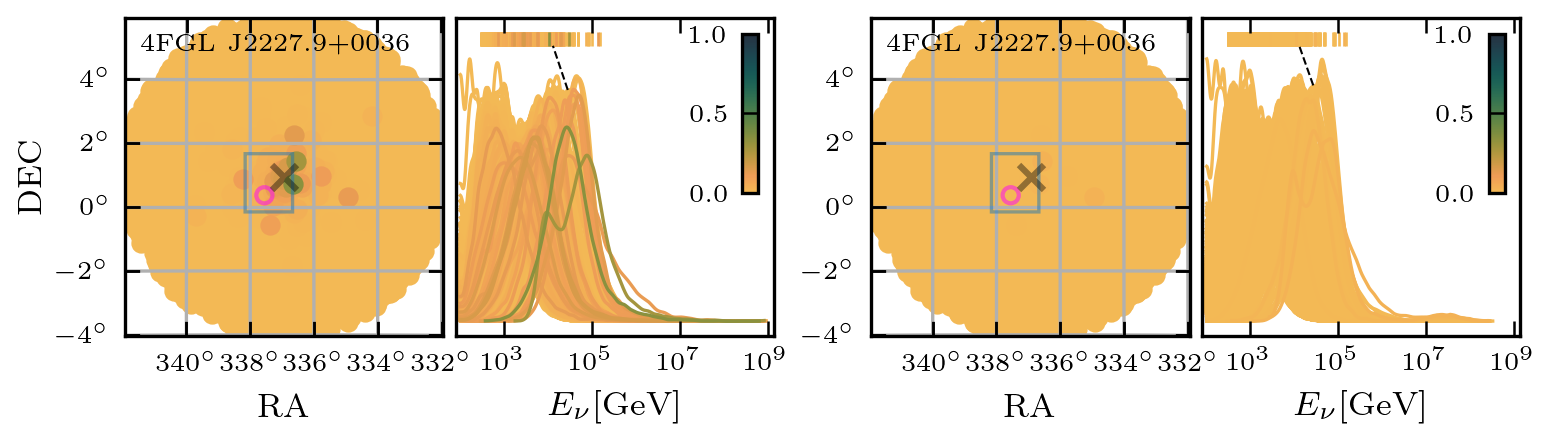}
  \caption{Comparison of power-law fits (left double column) to Fit \#2 (right double column). See Fig.~\ref{fig:app_roi_plots} for details.}
  \label{fig:app_roi_plots_pl_pgamma}
\end{figure}

\section{Prior influence}
\label{app:priors}
\restartappendixnumbering
We investigate the influence of the luminosity prior on the results, repeating fits
of representative cases (no event, one high-energy event, and multiple high-energy events from a point source)
with different luminosity priors. In particular, we set $\mu_L$ of a lognormal prior such that 0.1 and 1.0 events are expected,
and change the width of the prior to 4.0 and 6.0, leading to four different tested prior distributions per source.
Resulting posterior distributions of \Nex{}, \Lsrc{}, and \gammaps{}, together with summary statistics
on \Nex{} are shown in Fig.~\ref{app:prior_influence}.
\renewcommand{\thefigure}{B\arabic{figure}}
\begin{figure}[!h]
    \centering
    \includegraphics[width=\textwidth]{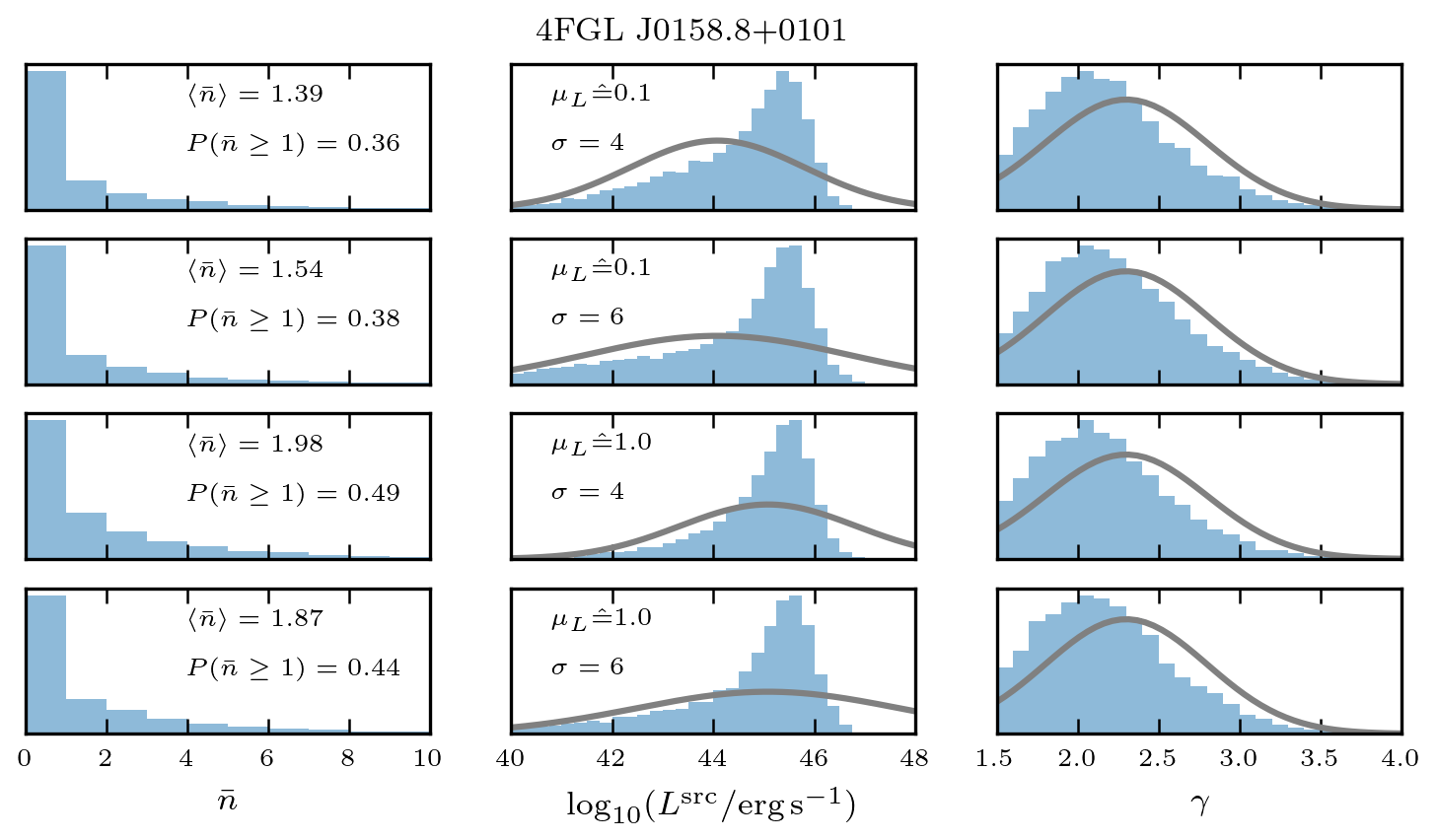}
    \includegraphics[width=\textwidth]{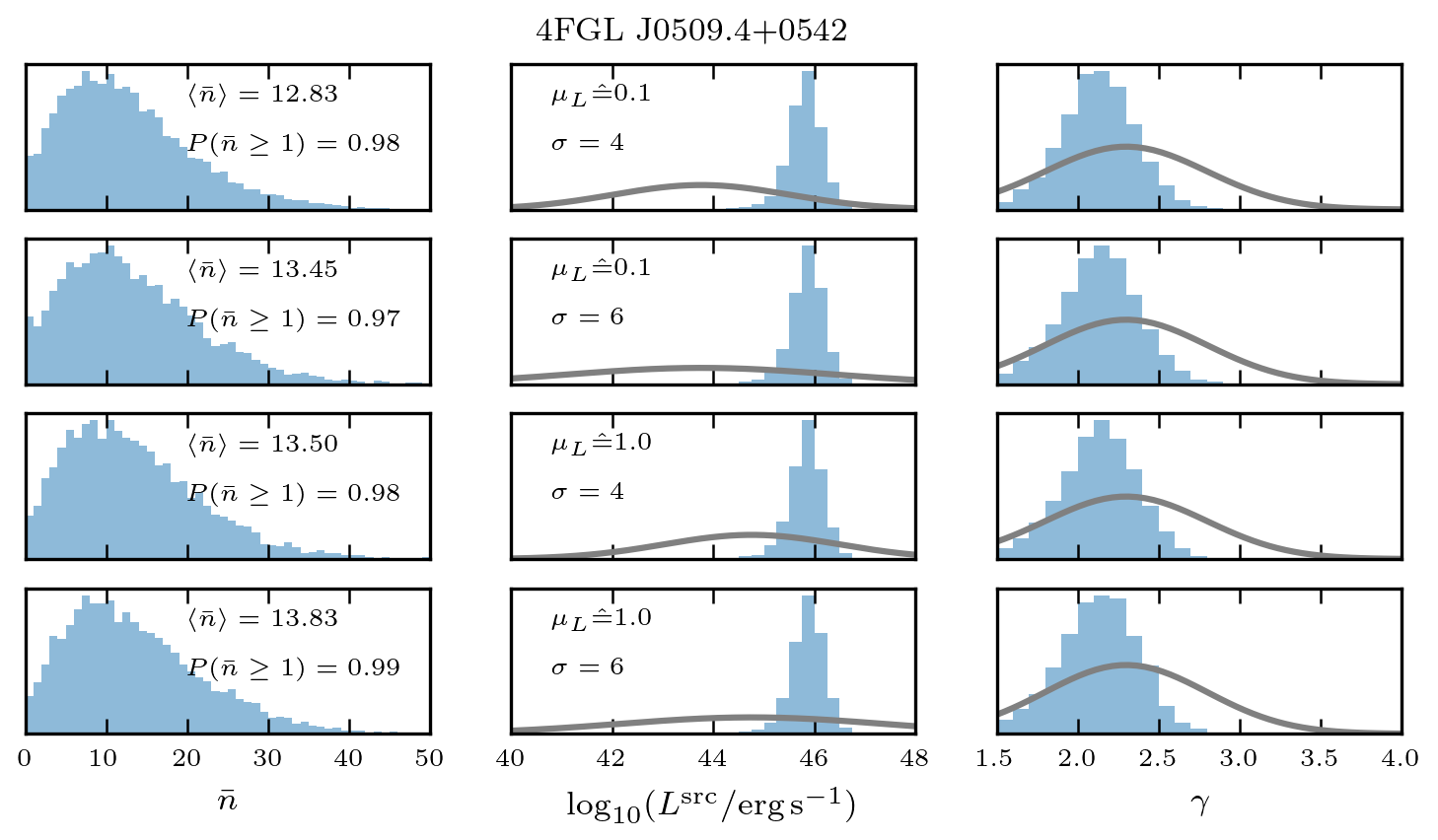}
    \caption{Prior dependency tests on fit results. }
    \label{app:prior_influence}
\end{figure}
\renewcommand{\thefigure}{B\arabic{figure} (Cont.)}
\addtocounter{figure}{-1}
\begin{figure}[!h]
    \centering
    \includegraphics[width=\textwidth]{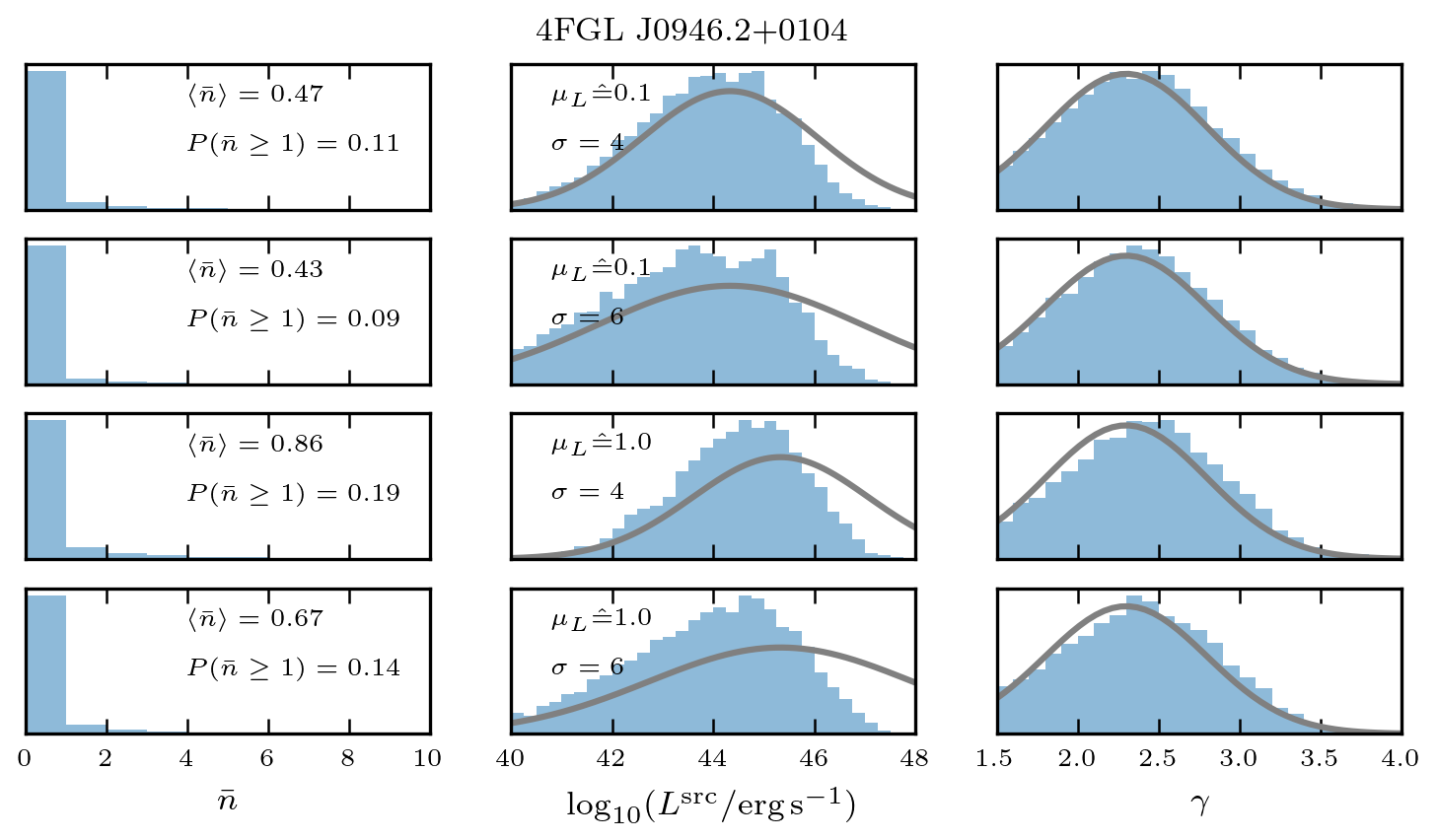}
    \includegraphics[width=\textwidth]{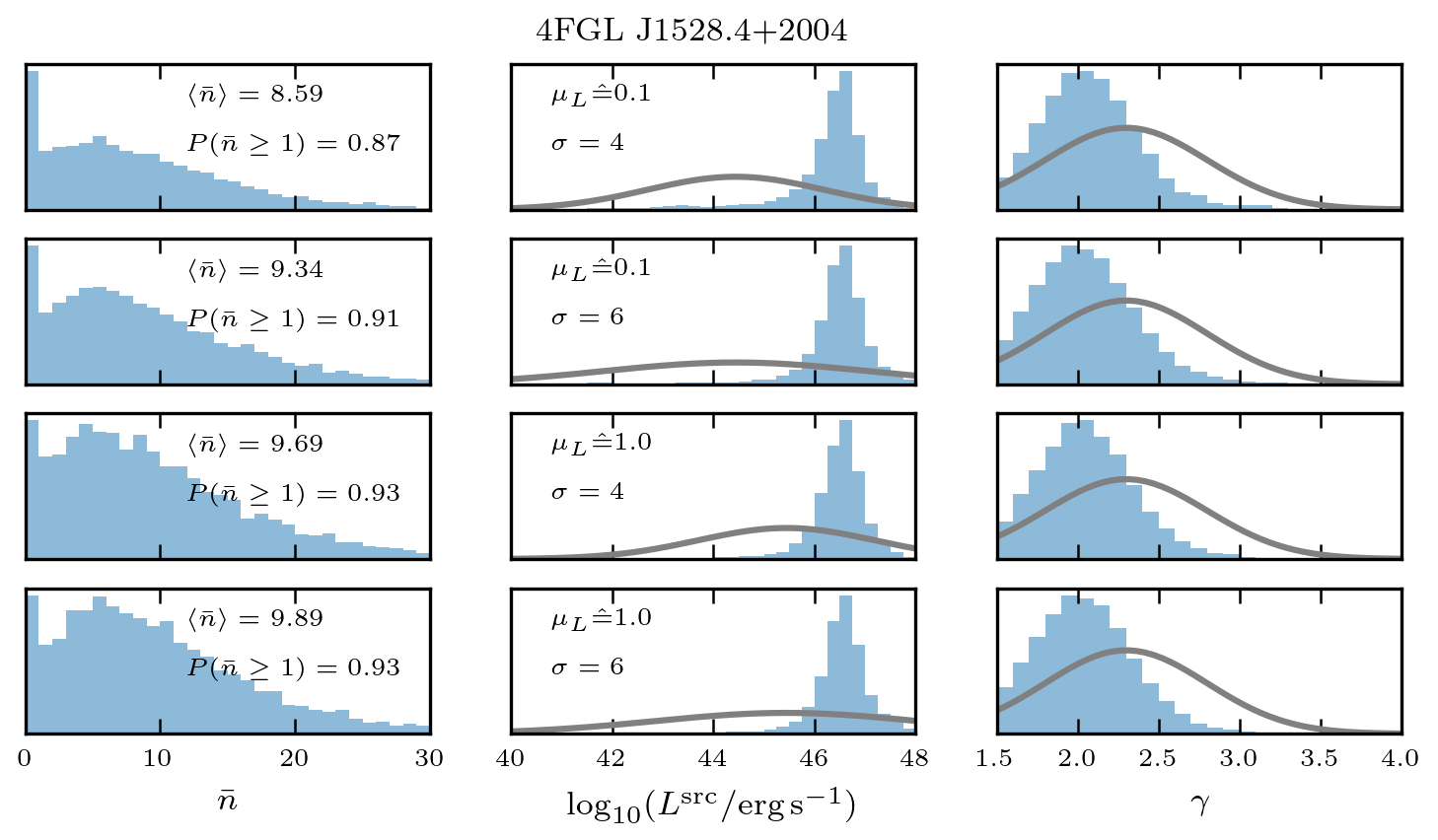}
    \caption{}
\end{figure}
\renewcommand{\thefigure}{B\arabic{figure}}

\bibliography{hierarchical_nu}{} \bibliographystyle{aasjournal}

\end{document}